\documentclass[12pt]{article}
\usepackage{scicite}
\usepackage{times}
\usepackage{amsmath}    
\usepackage{color}
\usepackage[10pt]{moresize}
\usepackage{caption}

\DeclareCaptionLabelFormat{adja-page}{\hrulefill\\#1 #2 \emph{(previous page)}}
\usepackage{subfig}
\usepackage{graphics,graphicx}
\usepackage{amsfonts}
\usepackage{amssymb}
\usepackage{amscd}
\usepackage{amsmath}
\usepackage{enumerate}
\usepackage{epsfig}
\usepackage{epstopdf}
\usepackage{dcolumn}
\usepackage{bm}
\usepackage{amsthm}
\usepackage{amsfonts}
\usepackage{color} 
\usepackage[usenames,dvipsnames]{xcolor}
\usepackage{amsmath}
\usepackage{enumerate}
\usepackage{bbm}
\usepackage{gensymb}
\usepackage{titling}
\usepackage{algorithm}
\usepackage{algorithmic}

\usepackage[figuresright]{rotating}

\usepackage[colorlinks=true,citecolor=blue,urlcolor=black]{hyperref}

\usepackage[normalem]{ulem} 
\usepackage{gensymb} 

\usepackage{mathrsfs}

\newenvironment{sciabstract}{%
	\begin{quote} \bf}
	{\end{quote}}

\title{Super-resolution imaging based on active optical intensity interferometry}

\author
{Lu-Chuan Liu$^{1,2,3*}$, Cheng Wu$^{1,2,3*}$, Wei Li$^{1,4,5*}$, Yu-Ao Chen$^{1,2,3}$,\\
 Frank Wilczek$^{6,7,8,9,10}$, Xiao-Peng Shao$^{4,5}$, Feihu Xu$^{1,2,3}$,\\
Qiang Zhang$^{1,2,3,11}$ and Jian-Wei Pan$^{1,2,3}$\\
	\\
	\normalsize{$^1$Hefei National Research Center for Physical Sciences at the Microscale and School of Physical Sciences,}\\
	\normalsize{University of Science and Technology of China, Hefei, 230026, China}\\
	\normalsize{$^2$Shanghai Research Center for Quantum Science and CAS Center for }
	\\
	\normalsize{Excellence in Quantum Information and Quantum Physics,}
	\\
	\normalsize{University of Science and Technology of China, Shanghai, 201315, China}\\
	\normalsize{$^3$Hefei National Laboratory, University of Science and Technology of China, Hefei, 230026, China}\\
	\normalsize{$^4$Hangzhou Institute of Technology, Xidian University, Hangzhou, 311200, China}\\
	\normalsize{$^5$School of Optoelectronic Engineering, Xidian University, Xi’an, 710071, China}\\
	\normalsize{$^6$Center for Theoretical Physics, MIT, Cambridge, MA 02139, USA}\\
	\normalsize{$^7$T. D. Lee Institute, Shanghai Jiao Tong University, Shanghai, 200240, China}\\
	\normalsize{$^8$Wilczek Quantum Center, School of Physics and Astronomy,}
	\\
	\normalsize{Shanghai Jiao Tong University, Shanghai, 200240, China}\\
	\normalsize{$^9$Department of Physics, Stockholm University, Stockholm, SE-106 91, Sweden}\\
	\normalsize{$^{10}$Department of Physics, Arizona State University, Tempe, AZ 85287, USA}\\	
	\normalsize{$^{11}$Key Laboratory of Space Active Opto-Electronic Technology, Shanghai Institute of Technical Physics,}
	\\
	\normalsize{Chinese Academy of Sciences, Shanghai, 200083, China}\\
	\normalsize{$^\ast$ These authors contributed equally to this work.}\\
}

\date{}

\begin{document}

\baselineskip24pt


\maketitle 

\begin{sciabstract}

Long baseline diffraction-limited optical aperture synthesis technology by interferometry plays an important role in scientific study and practical application. In contrast to amplitude (phase) interferometry, intensity interferometry --- which exploits the quantum nature of light to measure the photon bunching effect in thermal light --- is robust against atmospheric turbulence and optical defects. However, a thermal light source typically has a significant divergence angle and a low average photon number per mode, forestalling the applicability over long ranges. Here, we propose and demonstrate active intensity interferometry for super-resolution imaging over the kilometer range. Our scheme exploits phase-independent multiple laser emitters to produce the thermal illumination and uses an elaborate computational algorithm to reconstruct the image. In outdoor environments, we image two-dimension millimeter-level targets over 1.36 kilometers at a resolution of 14 times the diffraction limit of a single telescope. High-resolution optical imaging and sensing are anticipated by applying long-baseline active intensity interferometry in general branches of physics and metrology.

\end{sciabstract}

\section*{Introduction}

Hanbury Brown and Twiss (HBT) first proposed the seminal idea of high-angle resolution intensity interferometry for the measurement of stellar diameter~\cite{brown1956test,brown1957interferometry,brown1974intensity}, which is a technique based on the measurement of temporal correlations of arrival time between photons recorded in different detectors. The intensity interferometry does not require a high-precision phase stabilization system, and it is robust against optical imperfections and atmospheric turbulence. Hence, it is much easier to implement than amplitude (phase) interferometry. In astrophysics, the forthcoming air Cherenkov telescope arrays, which consist of almost 100 telescopes, are the current largest optical intensity interference project~\cite{cta2011design,abeysekara2020demonstration}. Following its start in astronomy, intensity interferometry has been developed as a versatile lab tool in general research fields, including the probe of interactions in high-energy particle physics~\cite{boal1990intensity}, studying the photon propagation in nonlinear media~\cite{bromberg2010hanbury} and curved space~\cite{schultheiss2016hanbury}, measurements of quantum correlations in ultracold bosonic and fermionic systems~\cite{Henny1999,jeltes2007comparison,dall2013ideal}, and identification of single photon source~\cite{eisaman2011invited}. 

Although intensity interferometry has been widely implemented in passive imaging scenes~\cite{dravins2012stellar,malvimat2014intensity}, it remains challenging to apply to the field of active imaging, i.e., light detection and ranging (LiDAR). Benefiting from active illumination, LiDAR has emerged as a powerful tool to image non-self-luminous targets~\cite{behroozpour2017lidar,rapp2020advances}. To probe high resolution and resist atmospheric turbulence, active intensity interferometry has emerges as a great candidate. However, the lack of collimated a narrow-spectrum thermal light source and a robust imaging retrieval algorithm imposes practical difficulties in its applications in active super-resolution imaging. 

Intensity interference originates from the quantum nature of thermal light sources, which can be understood as the photon bunching effect~\cite{brown1974intensity,glauber1963photon}. The shape of the bunching peak measured by the intensity interferometer is squared proportional to the Fourier transform of the shape of the source~\cite{van1934wahrscheinliche,zernike1938concept}. In experiments, intensity interference is obtained by cross-correlating intensity fluctuations measured in different pairs of telescopes to achieve aperture synthesis and yield resolution enhancement. In general, unlike lasers, thermal light sources cannot simultaneously meet the requirements of intense light power, narrow spectra and small divergence angles. A better alternative solution involves utilizing a composite approach integrating a laser source with a spatial-temporal modulation device, such as a rotating ground glass~\cite{valencia2005two,ferri2005high,katz2014non}, a spatial light modulator~\cite{shapiro2008computational}, and a projector~\cite{sun20133d} to synthesize a pseudo-thermal light source. However, these pseudothermal light sources are still incompetent for long-range illumination.

In this letter, we show that a phase-independent multiple laser emitter array can act as a pseudothermal light source, satisfying both the thermal nature of the source and the requirement of long-range illumination for intensity-interferometry imaging. In the experiment, we illuminate the targets with eight independent laser emitters modulated by the atmosphere and receive the back-reflected photons using two telescopes with configurable transverse range. For image retrieval, we develop an elaborate data preprocessing technique equipped with a modified phase retrieval algorithm, each tailored to adapt the intensity interferometric data. Our algorithm copes with sparse sampling in the interferometric plane, the nonlinearity of the imaging model, and the loss of Fourier phase information. Together, we experimentally realize super-resolution imaging over 1.36~km in an outdoor urban environment. The imaging resolution is 3 mm, which is 14 times higher than the diffraction limit of a single telescope.

\section*{Theory of active optical intensity interferometry}
\label{Theory}

An optical intensity interferometry in a classical imaging system typically necessitates that the imaging target to be a thermal light source to measure the non-shot-noise-based intensity variations of the two light detectors and compute their correlation. For ideal monochromatic polarized thermal light, Hanbury Brown and Twiss  showed that the second-order intensity correlation function is proportional to the square of the modulus of the Fourier transform of the intensity distribution on the target surface. We now express it in mathematical form. Suppose we have two light detectors $D_a$ and $D_b$, whose intensities measured at time $t$ are $I_a(t)$ and $I_b(t)$, respectively, and let $\textbf{k}_a$ and $\textbf{k}_b$ be the wave vectors from the target to the detectors $D_a$ and $D_b$, respectively. Then, according to the intensity interferometry theory~\cite{brown1974intensity}, the normalized intensity correlation function $c_{ab}^{(2)}$ can be expressed as
\begin{equation}
	c_{ab}^{(2)}=\frac{\langle\Delta I_a(t)\Delta I_b(t)\rangle}{\langle I_a(t) \rangle \langle I_b(t) \rangle}=|f(\textbf{k}_a-\textbf{k}_b)|^2,
	\label{idealhbt}
\end{equation}
where $\langle...\rangle$ is the time averaged notation, $\Delta I(t)\equiv I(t)-\langle I(t)\rangle$ and $f$ is the normalized Fourier function of the target surface intensity distribution $\rho(\textbf{r})$, that is
\begin{equation}
f(\Delta\textbf{k})=\frac{\int\rho(\textbf{r})\mathrm{e}^{-\mathrm{i}\Delta\textbf{k}\cdot\textbf{r}}\mathrm{d}\textbf{r}}{\int\rho(\textbf{r})\mathrm{d}\textbf{r}}.
	\label{normFourierTrans}
\end{equation}
Let $L$ be the distance from the target to the interferometer and $\lambda$ be the wavelength of light. 
Since $\Delta \textbf{k}$ and the projected interferometer baseline $\textbf{B}$ have a proportional relationship that is  $\Delta\textbf{k}=2\pi\textbf{B}/(\lambda L)$, super-resolution can be achieved by changing $\textbf{B}$ to cover a sufficiently large  area of the $u$ - $v$ plane to obtain enough information about the target in the Fourier domain.  The angular resolution of the intensity interferometer is about $\lambda/ B_{max}$ at the wavelength $\lambda$, which is larger than that of a single telescope~\cite{brown1974intensity}. 

\begin{figure}[ht]
	\centering
	\includegraphics[width=0.8 \linewidth]{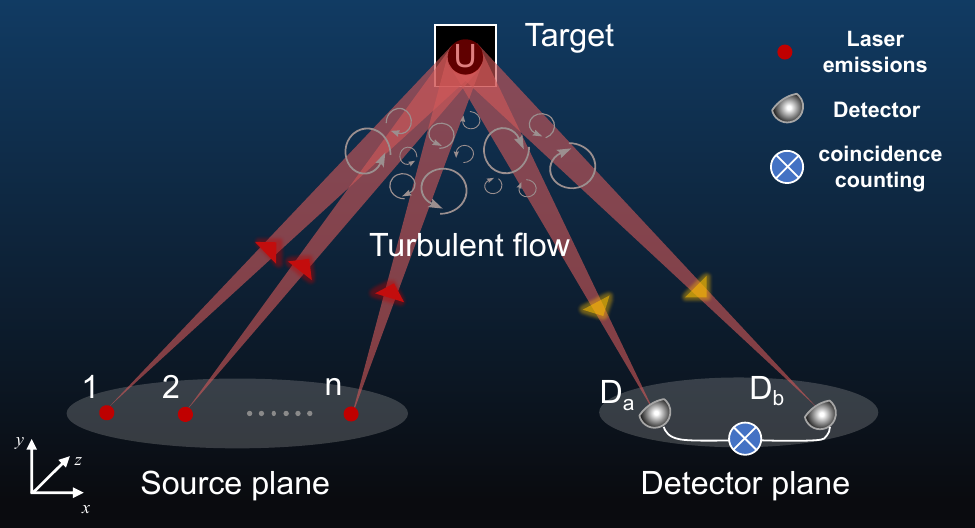}
	\caption{\textbf{Experimental schematic.} The active optical intensity interferometer consists of two parts: the source plane and the detection plane. The source plane contains $n$ ($n$ = 8 in the present experimental setup) laser emitters arranged at fixed equal intervals $d$ = 0.15~m. These laser emitters are generated from the same laser source and then evenly divided by beam splitters to irradiate 8-way light toward the target. The phase of each path of light is randomly modulated by passing different atmospheric turbulence to superpose pseudothermal illumination. The target is affixed to a rotating base, facing toward the source plane. The detection plane consists of two single photon avalanche diode (SPAD) detectors mounted on a one-dimensional linear displacement stage for obtaining changeable baselines. Coincidence count sets are gathered by cross-correlating intensity fluctuations from $D_a$ and $D_b$ to achieve the intensity correlation function $c_{ab}^{(2)}$.}
	\label{fig: forward model}
\end{figure}

However, if intensity interferometry is used for active imaging, it is hard to illuminate distant targets with poorly collimated thermal light sources. On the other hand, if we illuminate the target with well-collimated coherent light, such as a laser, intensity interference cannot be observed because the light detector cannot measure intensity signal fluctuations other than the shot noise, which is not correlated with the shot noise in another detector. 

To overcome the above difficulties, a basic idea is to superimpose a collection of phase-independent coherent lights to achieve pseudothermal illumination. The working principle of active intensity interferometry is schematized in Fig.~\ref{fig: forward model}. To simply verify the feasibility of applying super-resolution optical intensity interferometry to active imaging, as indicated in Fig.~\ref{fig: forward model}, we bypass the need for phase modulators by instead using atmospheric turbulence to randomly modulate the phases of multiple laser emitters. In the case of such non-ideal pseudothermal illumination, speckle-like noise occurs when measuring the intensity correlation function, so we propose a statistical optics theory (see supplementary materials), in which Eq.~\ref{idealhbt} needs to be modified to a form of the ensemble average, namely

\begin{equation}
	\langle c_{ab}^{(2)}\rangle_e=c_0+c_1|f(\textbf{k}_a-\textbf{k}_b)|^2,
	\label{activehbt}
\end{equation}

where $c_0$ and $c_1$ are the coefficients jointly determined by the light intensity, the autocorrelation coefficient of different emitters, and the optical memory effect~\cite{feng1988correlations,freund1988memory}. If we ignore the optical memory effect and consider a case where all laser emitters are symmetrically equivalent, then $c_0=c/n$ and $c_1=(n-1+c)/n$ are simple functions of the number of laser emitters $n$ and the autocorrelation coefficient $c$.

The above statistical linear relationship allows us to extract information about the target in the Fourier domain from the measurement of the intensity correlation function. Specifically, considering that the ensemble average expression predicts the average result of multiple parallel measurements, for a single shot measurement of $c_{ab}^{(2)}$, we can write $c_{ab}^{(2)}=\langle c_{ab}^{(2)}\rangle_e+\epsilon$, where $\epsilon$ is the speckle-like noise term that we wish to suppress as much as possible. We calculate an expression for the noise intensity (i.e., the standard deviation of $c_{ab}^{(2)}$) and find that it decreases to 0 at a rate of $O(1/\sqrt{n})$ as the number of laser emitters $n$ increases (see supplementary materials). This means that one way to suppress the experimental noise is to increase the number of laser emitters so that they are physically closer to thermal illumination. We also discuss the boundary cases of single ($n=1$) and infinite ($n=\infty$) laser emitters (fig.~\ref{fig: double slit different emissions}), and show that the former will lead to a very low signal-to-noise ratio (SNR), while the latter will give predictions that are consistent with classical HBT theory based on ideal thermal light sources.

\section*{Intensity Interferometry Imaging Setup}
\label{setup}

\begin{figure}[ht!]
	\centering
	\includegraphics[width=0.8 \linewidth]{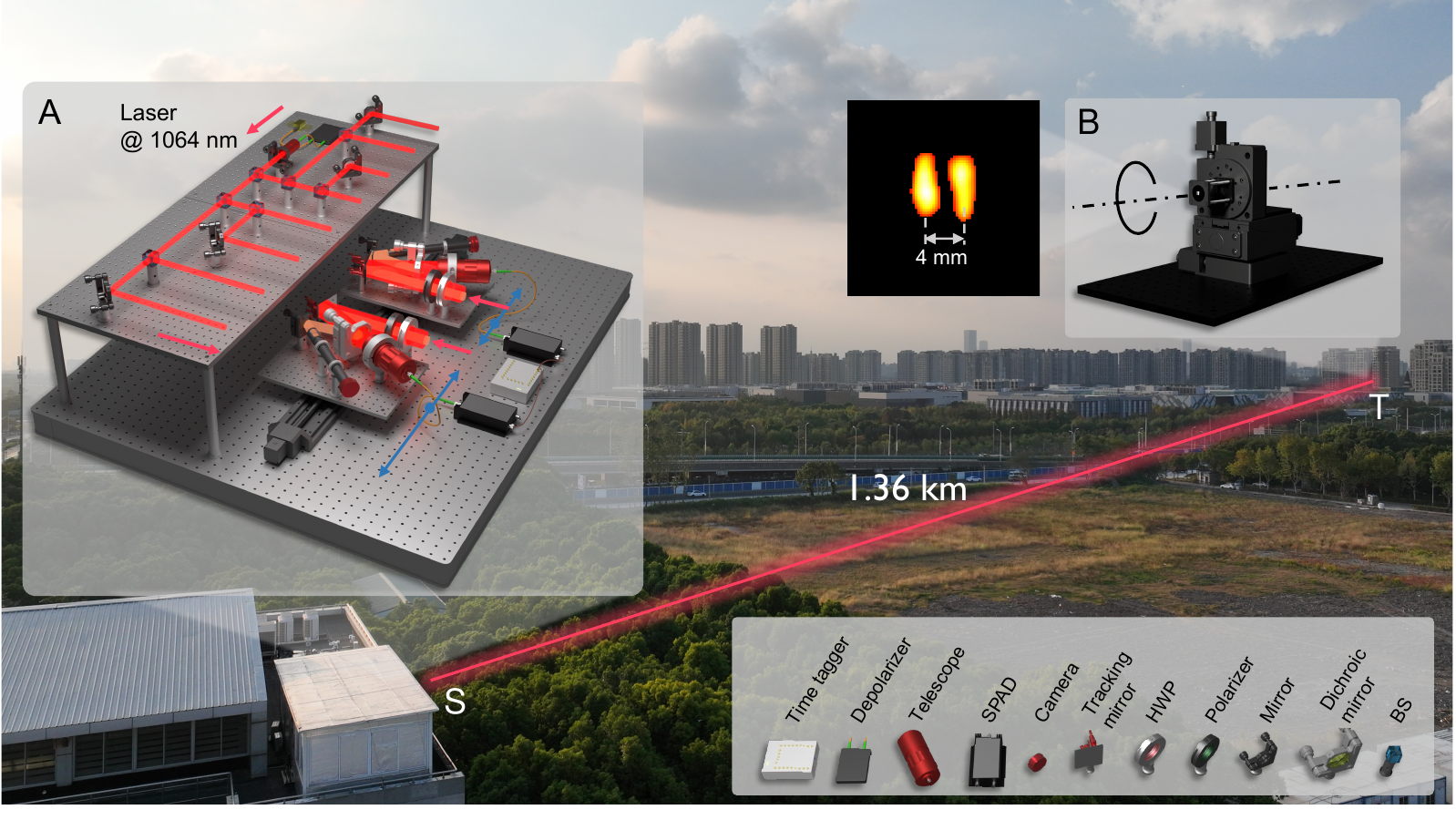}
	\caption{\textbf{Super-resolution imaging based on the intensity interferometer.} An aerial diagram of the imaging experiment over a 1.36-km free-space link, with the setup at S and the imaging target at T. (\textbf{A}) The optical setup of the imaging system, which consists primarily of a  continuous-wave (CW) laser emitter array and two receiving telescope systems for the intensity interferometer. In the laser emitter array system, the light is depolarized and coupled into free space, through a number of beam splitters (BSs), split into 8 beams. In the intensity interferometer,  two receiving telescope systems move with the translation stage to change the baseline in the range of 0.07-0.87~m. Both telescope systems are configured to receive left circularly polarized light through a quarter-wave plate (QWP) and a polarizer. The light collected by the receiving system is separated by the dichroic mirror, with the signal photon part being coupled into the multi-mode fiber by the telescope and guided to the single-photon avalanche diode (SPAD), and the background light part being received by the camera for feedback on the tracking mirror pointing.  The arrival time information of photons is recorded by the time-to-digital converter (TDC). (\textbf{B}) Schematic of the imaging target. The imaging target is put on the motorized rotation stage. (Insets) Imaging result of a double slit.}
	\label{fig:setup}
\end{figure}

We implement an intensity interferometry imaging experiment in an urban atmospheric environment 1.36 kilometers away. Figure ~\ref{fig:setup} shows an aerial view of the experiment configuration, with the imaging system placed at location S, facing the imaging target located at T over a distance of 1.36 km (determined by laser ranging measurement). The target is put on the 3-axis motorized rotation stages, as shown in Fig.~\ref{fig:setup}B. 

The active optical intensity interferometer consists of phase-independent multiple laser emitters and two identical receiving telescope systems, as shown in Fig.~\ref{fig:setup}A. A $\lambda$ = 1064~nm CW laser ($ \sim$100~mW) is depolarized by a single-mode optical fiber depolarizer to reduce the intensity fluctuation caused by different polarization reflectivities of the target. The light is coupled into free space by a collimator with a focal length of 40~mm and is divided into 8 beams by a beam splitter group with equal spatial separation and power. Eight beams are adjusted to point at the target. The size of each beam after 1.36~km is approximately 0.2~m, more than the size of the target. Since light is subject to phase fluctuations through atmospheric turbulence, the phase shifts of each beam are randomly modulated~\cite{fried1966optical,ridley2011measurements}. The distance between each beam at the laser emitter array is set to 0.15~m, more than the outer scale of atmospheric turbulence (the atmospheric coherence length typically measured 0.02$\sim$0.05 m), to ensure that each beam has an independent random phase shift. In the theoretical analysis, a collection of phase-independent multiple coherent light can be equivalent to a pseudothermal light source. 

Two identical receiver systems $D_a$ and $D_b$ are separately placed on two 0.4-m-long linear translation stages to change the baseline in the range of [0.07, 0.87]~m. The receiving system is designed for rotational symmetry, collecting circularly polarized light through a quarter-wave plate and a polarizer. A tracking mirror mounted on a piezo platform is used to adjust the receiving field of view (FOV), ensuring that the target remains within the FOV while the telescope system moves. A dichroic mirror is introduced spectrally to separate the collected light. The signal light with a wavelength of 1064~nm transmitting to the dichroic mirror is coupled to a 62.5-$\mu$m multi-mode optical fiber with a telescope that has a diameter of 42.5~mm and a focal length of 80~mm, guided to the silicon single-photon avalanche diode (SPAD). Other background light is received by the camera with a 300~mm focal length lens for feedback on the tracking mirror pointing. 

The arrival time information of photons at two SPADs is recorded using a time-to-digital converter (TDC) with a time resolution of 8~ps, from which the normalized intensity correlation function $c_{ab}^{(2)}$ can be calculated. The optical delay difference between the two receiving systems is controlled below 1~ns. Considering the optical coherence time of the atmosphere (typical value about 27~ms at 1064~nm, fig.~\ref{fig: g2_tau}), the time bin and the integration time are set to 2~ms and 2~s, respectively. The count rate of each SPAD is about $10^4$ photons per second and the SNR of coincidence counting measurement is about 10 (see supplementary materials for more details).

\section*{Image Reconstruction}

\subsection*{Data acquisition.}

In the measurement, we change the baseline distance between two receiving telescope systems from 0.07~m to 0.87~m per 0.04~m steps. When the baseline is set to a specific length, the target is rotated every $6^{\circ}$ in the range of 0 - $360^{\circ}$, and the intensity correlation function $c_{ab}^{(2)}$ is obtained from the coincidence counts detected by $D_a$ and $D_b$. A total of 60 $\times$ 21 = 1260 coincidence count sets are gathered, and the corresponding Fourier $(u,v)$ coverage is depicted in Fig.~\ref{fig: data processing}A. 

\subsection*{Reconstruction.}

\begin{figure}[ht!]
	\centering
	\includegraphics[width=0.8 \linewidth]{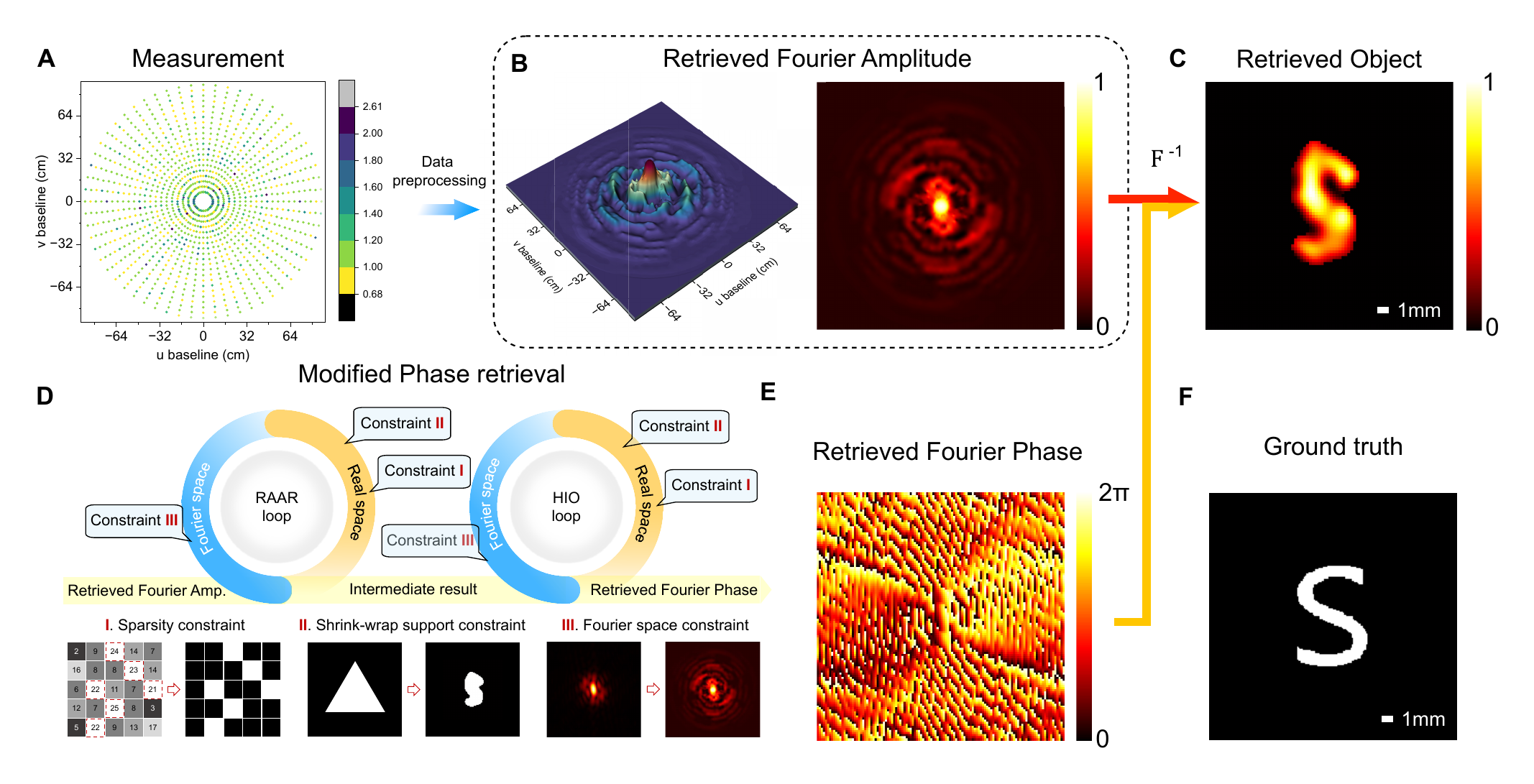}
	\caption{\textbf{Data processing. }(\textbf{A}) Intensity correlation function $c_{ab}^{(2)}$ measured at $\lambda$ = 1064 nm for the letter “S” (F) with a distance of 1.36 km. (\textbf{B}) Fourier magnitude derived from elaborated data preprocessing (see supplementary materials). The pseudo-colored surface represents the preprocessed target Fourier spectrum. After combining the retrieved Fourier magnitude (B) and retrieved Fourier phase (E), the inverse Fourier transform yields the final retrieved target (\textbf{C}). (\textbf{D}) The flowchart of our modified phase retrieval algorithm. As an input, the retrieved Fourier magnitude (B) is alternately projected and reflected in the Fourier and real domains using a cascaded RAAR and HIO strategy equipped with three types of constraints (I and II in real space and III in Fourier space). The real space constraint is a hybrid of an adaptive shrink-wrap (II) and a sparse constraint (I), while the Fourier space constraint is an update of the iterative intermediates as the retrieved Fourier magnitude (B) (III). (\textbf{E}) Retrieved Fourier phase from (B) via algorithm (D). (\textbf{F}) The ground truth target. Scale bar: 1 mm ((C), (F)).}
	\label{fig: data processing}
\end{figure}

Intensity interferometers do not directly provide images but intensity correlation function that is proportional to the squared-magnitude of the Fourier transform of the source brightness~\cite{van1934wahrscheinliche,zernike1938concept}. To recover an image, the phase of the Fourier transform must be determined in addition to its magnitude~\cite{gerchberg1972practical,fienup2013phase,shechtman2015phase}. In real observations, however, there remain three main obstacles for existing image reconstruction techniques. First, the limited number of telescopes and observations yields very sparse frequency coverage. Second, since the square root operation needs to be applied to the intensity correlation function to obtain the Fourier magnitude, the imaging model is non-linear and especially sensitive to noise near zero values. Third, deploying the active intensity interferometer outdoors has to face strong urban environment noise, and the imperfection comes from the mimic pseudothermal light source. Hence robust phase retrieval techniques are needed. The reconstruction pipeline is shown in Fig.~\ref{fig: data processing}.

To address the nonconvex and ill-conditioned inverse problem, elaborate data preprocessing is designed to extract the target Fourier magnitude from the raw measurements. Before entering the phase retrieval procedure, the raw intensity correlation function $c_{ab}^{(2)}$ should pass through five successive processes. The data in polar coordinates were first averaged with adjacent multiple angles, and then central low-frequency estimation, target Fourier magnitude estimation, and high-frequency noise suppression were performed. Finally, the data were converted into Cartesian coordinates and produced by two-dimensional Fourier interpolation to unveil the target Fourier magnitude. All data preprocessing is detailed in the supplementary materials.

After data preprocessing, we finally obtained the retrieved target Fourier magnitude $y =|F(\rho(\mathbf{r}))|$ (Fig.~\ref{fig: data processing}B). As another key ingredient in the reconstruction process, phase retrieval is designed to reconstruct an image from the Fourier-plane magnitude and object-plane constraint. Let $\hat{x} = \widehat{\rho(\mathbf{r})} \in \mathbb{C}^n$ be the target surface intensity distribution and $y \in \mathbb{R}_{+}^N$ be the acquired data of $|F \hat{x}|$. The phase retrieval problem is to find an approximate solution to the equation,
\begin{equation}
	y =|F x|+w, \quad x \in \mathbb{C}^n
	\label{phase_retrieval}
\end{equation}
where $w \in \mathbb{R}^N$ represents the unknown discrepancy between the data predicted according to the Fourier transform model $F$ and the actually acquired data. Mathematically, Eq.~\ref{phase_retrieval} can be modeled as
\begin{equation}
	\hat{x}=\arg \min _x \Phi(x)+\lambda R(x),
	\label{phase_retrieval_inv}
\end{equation}
where $\Phi(x) = 1 / 2\|y-|F x|\|_2^2$ is a data-fidelity regularizer that ensures consistency between the reconstructed result and measurements, and $R(x)$ is a regularizer that imposes a certain prior constraint. 

To solve the above problem (Eq.~\ref{phase_retrieval_inv}), we altered the iterative alternating projection approach \cite{gerchberg1972practical,fienup1978reconstruction,marchesini2007invited} which incorporates a cascaded phase retrieval procedure composed of Relaxed Averaged Alternating Reflections (RAAR) \cite{luke2004relaxed} and Hybrid Input and Output (HIO) algorithm \cite{fienup1978reconstruction,fienup2013phase} with multiple constraints (sparsity, positivity, shrink support, and oversampling) (Fig.~\ref{fig: data processing}D). Putting all together, the elaborate data preprocessing followed by the modified phase retrieval strategy explicitly improves image recovery performance and stability (Fig.~\ref{fig: exp results}, fig.~\ref{fig: Different PR}, and movie S1). More detailed mathematical treatments of the imaging process appear in the supplementary materials.

\section*{Imaging Results}

\begin{figure}[ht!]
	\centering
	\includegraphics[width=0.6 \linewidth]{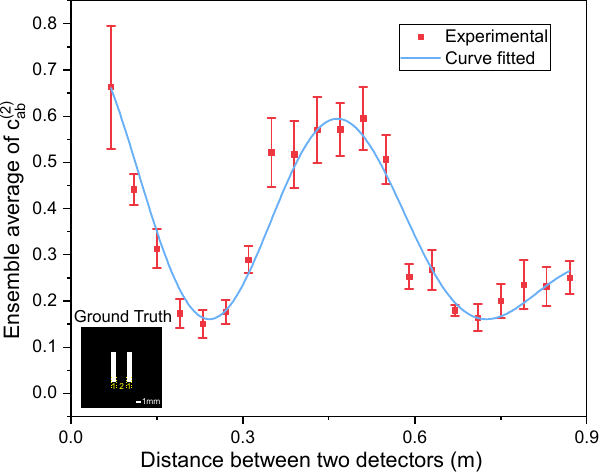}
	\caption{\textbf{Experimental results and fitting of a double slit in one dimension.} In the baseline direction, the two slits are both 1 mm wide, and the center distance between the two slits is 3 mm. We assume the spatial ergodicity of speckle~\cite{1990Looking} and use the average of the $c_{ab}^{(2)}$ measurements corresponding to 100 random minor changes in the target attitude as an ensemble average measurement (red dot). The blue curve is the least squares fitting result of Eq. (\ref{activehbt}) with fitting parameters $c_0=0.160$ and $c_1=0.625$.}
	\label{fig: 1D_double_slit}
\end{figure}

\begin{figure}[ht!]
	\centering
	\includegraphics[width=1 \linewidth]{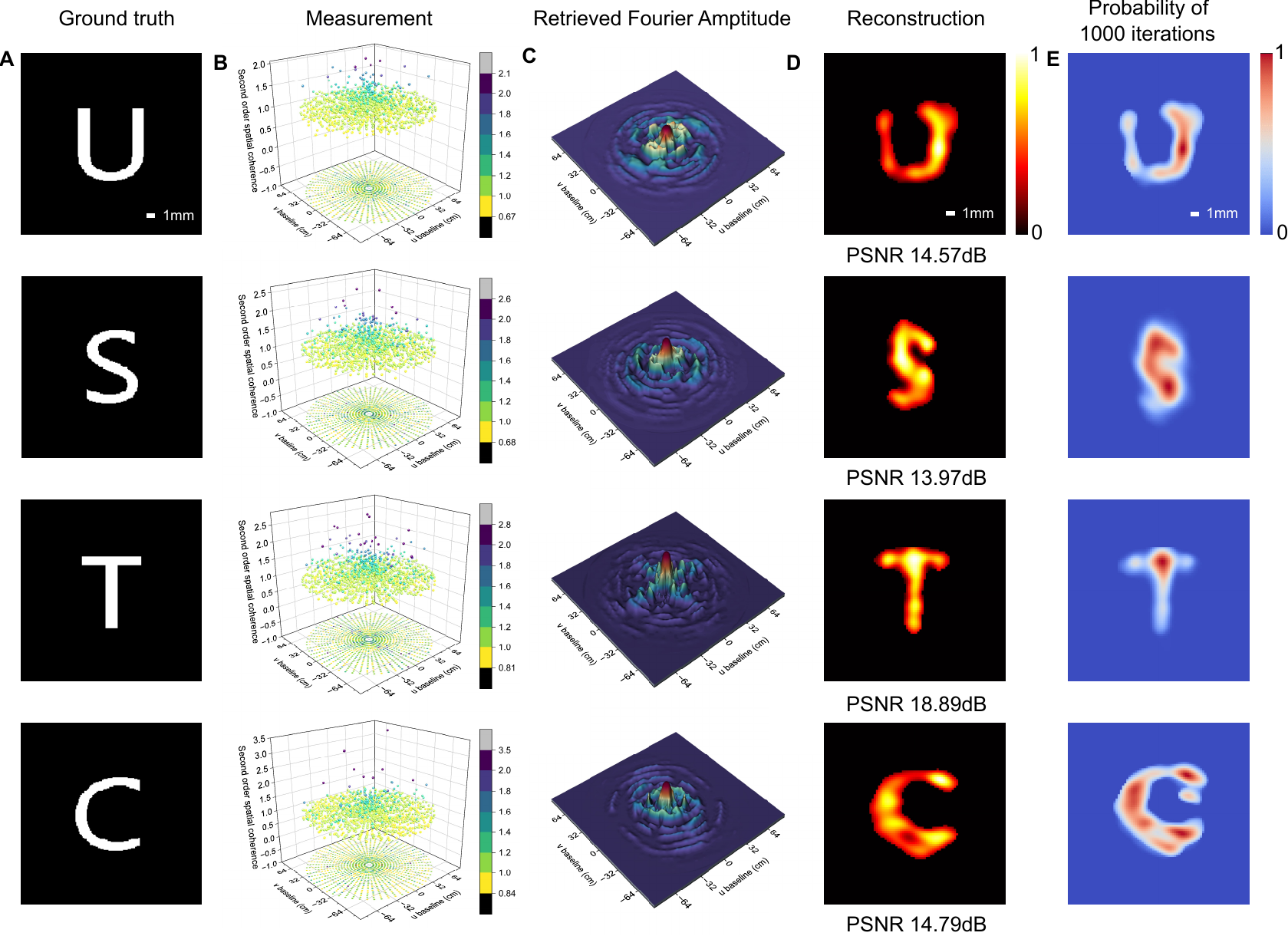}
	\caption{\textbf{Main experimental results.} The first column (\textbf{A}) shows the ground truth targets placed at a distance of 1.36 km. (\textbf{B}) Intensity correlation function $c_{ab}^{(2)}$ measured at $\lambda$ = 1064 nm for artificial targets listed in (A). Colored dots with 2D $x$-$y$ projection mark the observed sparse data points. (\textbf{C}) Retrieved Fourier magnitude $|F(\rho(\mathbf{r}))|$ from (B) via data preprocessing (detailed in the Fig.~\ref{fig: Flow chart} (E-K)). (\textbf{D}) Retrieved target images from (C) using the modified phase retrieval algorithm (see Fig.~\ref{fig: data processing}D and supplementary materials). The fifth column (\textbf{E}) shows the retrieval probability map with 1000 phase retrieval algorithm trials. Scale bars: 1 mm ((A), (D), (E)).}
	\label{fig: exp results}
\end{figure}

We demonstrate the effectiveness of the proposed framework with both synthetic and experimental data. The resolution of our system is first confirmed by imaging a double slit. As shown in Fig.~\ref{fig: 1D_double_slit}, the result clearly shows that an image with a 3 mm cross-range resolution at approximately 1.36 km can be achieved, which is approximately 42.5 mm / 3 mm $\approx$ 14 times resolution enhancement over the single receiving aperture ($\approx 42.5 $ mm). The theoretical resolution predicted by the Rayleigh diffraction limit is $1.22\lambda L/B \approx 2.03$ mm, where $\lambda$, $L$, and $B$ represent the illumination wavelength, imaging distance and the size of the synthetic aperture, respectively. 

We produce super-resolution imaging results in Fig.~\ref{fig: exp results}. The targets (Fig.~\ref{fig: exp results}A) were crafted from surface-blackened aluminum sheets, hollowed out to create the letters "USTC", which are covered with retroreflective sheeting. The size of each individual letter is 8 mm by 9 mm, and the width of the character is 1.5 mm. The measured intensity correlation points at sampled frequencies are rebinned to match the grid of frequels with a size of 512 $\times$ 512 points. After applying the special data preprocessing pipeline (see fig.~\ref{fig: Flow chart} and supplementary materials), the targets' Fourier magnitude $|F(\rho(\mathbf{r}))|$ can be retrieved (Fig.~\ref{fig: exp results}C) and then used as input for the modified phase retrieval to obtain the two-dimensional target images (Fig.~\ref{fig: exp results}D). Shape reconstructions with PSNRs of 14 $\sim$ 18 dB demonstrate the feasibility of the proposed method. A superposition of 1000 individual trials of phase retrieval results is shown in Fig.~\ref{fig: exp results}E, which clearly demonstrates high image restoration fidelity and stability. Potential image quality improvement may be achieved by increasing the number of laser emitters $n$, the capture frames (fig.~\ref{fig: Different emissions vs acquisition frames}) as well as the sampling rate in the interferometric plane (figs.~\ref{fig: Different emissions vs acquisition frames} and ~\ref{fig: complex targets}).

\section*{Discussions}

In summary, we have shown active intensity interferometry for super-resolution imaging over the kilometer range. 
In the experiment, phase-independent multiple laser emitters based on the phase fluctuations of atmosphere turbulence can act as an elegant pseudothermal light source to illuminate long-range objects, but the exposure time of imaging is limited to atmospheric coherence time. With the introduction of an active phase modulation scheme and a confocal receiving-emitting telescope array, high-speed active intensity interferometry imaging is an interesting avenue for future work.
On the other hand, higher-fidelity reconstruction is anticipated by applying more advanced methods, such as the optimization strategy and deep learning modules. Furthermore, by using different speckle interferometry analysis techniques, such as phase closure analysis and amplitude interference, additional phase information may be obtained from the data in some cases, reducing the ambiguity of recovery and further improving the accuracy and fidelity of phase retrieval. In the future, efforts are being devoted to joint measuring the first and second-order coherence of light to uncover unrivaled imaging resolution. Further combining the active intensity interferometry with deep space exploration and microscopy will open a new path towards super-resolution imaging from macroscopic scale to microscopic scale. 

\bibliographystyle{Science}
\bibliography{HBT}

\section*{Acknowledgments}
We thank Zhen-Tao Liu and Shen-Sheng Han from Shanghai Institute of Optics and Fine Mechanics for their suggestions on this work. 

\paragraph*{Funding:}This research was supported by National Natural Science Foundation of China (grants T2125010 and 12204466); Natural Science Foundation of Shanghai (grant 22ZR1468200); Anhui Initiative in Quantum Information Technologies (grant AHY010100); Shanghai Municipal Science and Technology Major Project (grant 2019SHZDZX01); Innovation Programme for Quantum Science and Technology (grants 2021ZD0300100 and 2021ZD0300300). Y.-A. C., F.-H. X. and Q. Z. were supported by the XPLORER prize from New Corner Stone Science Foundation. F. W. was supported by the U.S. Department of Energy under grant Contract Number DE-SC0012567, by the European Research Council under grant 742104, and by the Swedish Research Council under Contract No. 335-2014-7424. 

\paragraph*{Author contributions:}Q. Z. and J.-W. P. conceived the research. L.-C. L. carried out the work of the theoretical derivation. L.-C. L. and C. W. built the experimental setup and collected the experimental data. W. L. worked on experimental simulations and developed the image reconstruction algorithms. All authors contributed to data processing, data analysis, and writing of the manuscript. 

\paragraph*{Competing interests:}The authors declare no competing interests. 

\paragraph*{Data and materials availability:}The measured intensity correlation data and the processing code supporting the findings of this study will be available online upon publication. Additional data and code are available from the corresponding authors upon request.

\section*{Supplementary Materials}
Materials and Methods\\
Figs. S1 to S6\\
References \textit{(36-42)}\\
Movie S1\\

\clearpage

\title{Supplementary Materials for Super-resolution imaging based on active optical intensity interferometry}

\author{Lu-Chuan Liu$^{1,2,3*}$, Cheng Wu$^{1,2,3*}$, Wei Li$^{1,4,5*}$, Yu-Ao Chen$^{1,2,3}$,\\
 Frank Wilczek$^{6,7,8,9,10}$, Xiao-Peng Shao$^{4,5}$, Feihu Xu$^{1,2,3}$,\\
Qiang Zhang$^{1,2,3,11}$ and Jian-Wei Pan$^{1,2,3}$}

\maketitle

\noindent\textbf{The PDF file includes:}

Materials and Methods

Figs. S1 to S6

References \textit{(36-42)}

\noindent\textbf{Other Supplementary material for this manuscript includes the following:}

Movie S1

\section*{Materials $\textbf{\&}$ Methods}

\section{Simulations.} 
\label{methods:simuations}

To gain further insight into the mechanism of active intensity interferometry, we performed numerical simulations of the whole process. The entire imaging processing and simulations are presented in Fig.~\ref{fig: Flow chart}, with the simulation process, data preprocessing and retrieval process indicated as red, orange, and blue backgrounds, respectively. To simulate the HBT sampling measurement, a target (Fig.~\ref{fig: Flow chart}A) is first transformed into the Fourier domain to achieve its Fourier magnitude (Fig.~\ref{fig: Flow chart}B). Then, the Fourier magnitude is sampled corresponding to the experimental measurement setup (see HBT imaging setup, data acquisition and Fig.~\ref{fig: Flow chart}C). The sparse Fourier sampling is then remapped using the linear transform to obtain the HBT sampling data i.e., the intensity correlation function $c_{ab}^{(2)}$ (Fig.~\ref{fig: Flow chart}D). The sampled $c_{ab}^{(2)}$ data were first converted into Polar coordinates (Fig.~\ref{fig: Flow chart}E) to simulate the real experiment acquisition. 

We describe our HBT simulation procedure in Algorithm \ref{alg:HBT simulation}. Please see Section \ref{Theory Part III} for a detailed explanation of the simulation process.

\begin{algorithm}
        \renewcommand{\algorithmicendfor}{\textbf{end}}
        \caption{HBT sampling simulation}\label{alg:HBT simulation}
        {\bf Input:} 
        $f$ (normalized complex Fourier spectrum), $n$ (number of lightning emitters), $c$ (autocorrelation coefficient)\\
        {\bf Step}
        \begin{algorithmic}[1]
        \STATE Initialize $X$ \COMMENT{complex array of length $n$}
        \FOR{$i$ from 1 to $n$} 
        \STATE $\left(x_1, y_1, x_2, y_2\right) \leftarrow$ Sampling 4 random real numbers from distribution $N\left(0, \sqrt{\frac{1}{2}}\right)$
        \STATE $\left(z_1, z_2\right) \leftarrow\left(x_1+j y_1, x_2+j y_2\right)$
        \STATE $X[i] \leftarrow z_2^*\left(\sqrt{1-|f(\textbf{k}_a-\textbf{k}_b)|^2}z_1+f(\textbf{k}_a-\textbf{k}_b)z_2\right)$
        \ENDFOR
        \RETURN $\left(\left|\sum_{i=1}^n X[i]\right|^2+(c-1) \sum_{i=1}^n|X[i]|^2\right) / n^2$
        \end{algorithmic}
\end{algorithm}

\newcounter{Sfigure}
\setcounter{Sfigure}{1}
\renewcommand{\thefigure}{S\arabic{Sfigure}}

\begin{figure}[ht!]
	\centering
	\includegraphics[width=1 \linewidth]{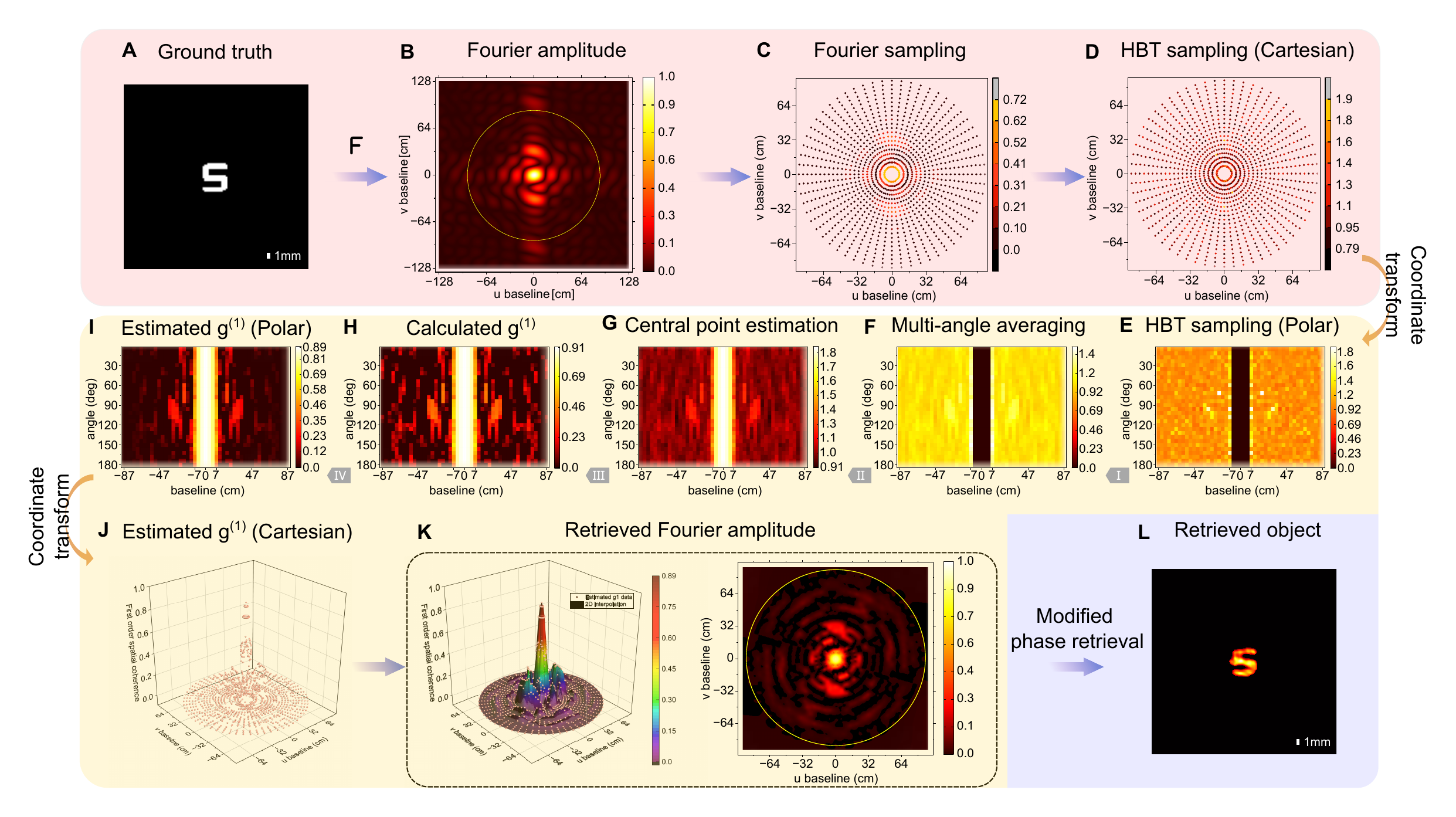}
	\caption{\textbf{The flow chart of the simulation and data processing pipeline.} The simulation, data preprocessing, and image reconstruction modules are represented by the red, orange, and blue backdrops, respectively. In the simulation module, the ground truth target (\textbf{A}) was first converted into the Fourier domain to obtain its Fourier magnitude (\textbf{B}), where the yellow circle denotes the current experimental setup's spatial frequency upper limit. (\textbf{C}) Fourier domain sampling also serves as a benchmark for the Fourier magnitude acquired by the ideal imaging system. A unitary matrix transformation was used to re-sample each spatial frequency for HBT sampling (\textbf{D}). (Section \ref{methods:simuations}) (\textbf{E}) HBT sampling in Polar coordinates. To predict the central low-frequency information, the central black region chosen at [-4 -2 0 2 4] (unit: cm) was preserved (E). In the data preprocessing module, the HBT sampling was first averaged with five adjacent angles and then row-by-row interpolated to obtain the central frequency estimation (\textbf{G}). Using Eq.~\eqref{activehbt}, one can calculate the $F|(\rho(\mathbf{r}))|$ (\textbf{H}) from (G). A Lamp-rank filter is then applied to eliminate high-frequency oscillation to obtain (\textbf{I}). The estimated $F|(\rho(\mathbf{r}))|$ data (\textbf{J}) were transformed into Cartesian coordinates and then 2-D interpolated to obtain the retrieved Fourier magnitude (\textbf{K}) with the yellow circle representing the maximum baseline region. In the image reconstruction module, target (\textbf{L}) is recovered from (K) via the proposed phase retrieval technique (Section \ref{methods:image processing} and Fig.~\ref{fig: data processing}D). Scale bars: 1 mm ((A), (L)).}
	\label{fig: Flow chart}
\addtocounter{Sfigure}{1}
\end{figure}

\section{Computational retrieval method.}

\subsection*{Image processing.}
\label{methods:image processing}

As described in the main text, to extract the target information from the measured intensity correlation function $c_{ab}^{(2)}$, we have to face three main challenges. First, the interferometric sampling remains very sparse (for the 1260 experiment data points fall in 512 $\times$ 512 grids, leading to a 0.48$\%$ sampling rate) (see Fig.~\ref{fig: data processing}B), and the central crucial low-frequency region (frequency components lie under the 0.07~m circle) is not measured due to the physical limitations of the distance between two telescopes. Second, to extract the target Fourier magnitude $|F(\rho(\mathbf{r}))| = \sqrt{c_1}|f(\textbf{k}_a-\textbf{k}_b)|$ with the Fourier transform operator defined as $F(\cdot)$ from $c_{ab}^{(2)}$, we need to calculate Eq.~\ref{activehbt}, which makes the model nonlinearity and noise sensitive to the close to zero values caused by the square root operation. Third, recovering the missing Fourier phase part is difficult due to the high background noise originating from the outdoor environment and source imperfections. 

To address this problem, the raw captured intensity correlation function $c_{ab}^{(2)}$ was first subtracted from the background and then polarization elimination was performed. The central symmetric average and adjacent five angle average were employed to make the data much approach to $\langle c_{ab}^{(2)}\rangle_e$, which obeys the Gaussian process (Section \ref{Theory Part I}). The central null part is preserved for low-frequency component estimation (Fig.~\ref{fig: Flow chart}E). Virtual frequency components at the positions of [-4 -2 0 2 4] cm were predicted by summing and averaging after 1-D cubic spline interpolation for each angle (Fig.~\ref{fig: Flow chart}F). The estimated target Fourier magnitude $|F(\rho(\mathbf{r}))|$ (Fig.~\ref{fig: Flow chart}G) was calculated using the Eq.~\ref{activehbt} ($c_0$ = 0.06 in Fig.~\ref{fig: exp results} and 0.15 in Fig.~\ref{fig: double slit different emissions}) and then smoothed by a Ramp-Lank filter to remove high-frequency oscillations (Fig.~\ref{fig: Flow chart}H). The filtered data were transformed into Cartesian coordinates and interpolated in two dimensions to fill the voids in the sparse sampled spatial frequencies to obtain the target Fourier magnitude. The resulting Fourier magnitude was oversampled with one-fold zero values, and a two-dimensional Hanning window was applied before entering the modified phase retrieval algorithm.

\subsection*{Retrieval algorithm.} 
\label{methods: retrieval}

As described in the Reconstruction section, to solve the highly ill-posed phase retrieval problem (Eq.~\ref{phase_retrieval_inv}), we leverage nonconvex optimization methods inspired by the alternating projections and reflections schemes~\cite{gerchberg1972practical,fienup1978reconstruction,fienup1986phase,marchesini2007invited}. Apart from normalization factors, we denote the target with reflectivity $\rho(\mathbf{r})$, $\mathbf{r}$ is the coordinates in the object (or real) space, and the corresponding Fourier Transform $\tilde{\rho}(\mathbf{k}) = F(\rho(\mathbf{r}))$, with $\mathbf{k}$ representing the coordinate in the Fourier (or Reciprocal) space. Four operators linking two sets $\mathbf{S}$ (support) and $\mathbf{M}$ (modulus) are introduced. Given a support set $\mathbf{S}$, the support projection operator $\mathbf{P}_{\mathbf{S}}$ involves setting to 0 the components outside the support while leaving the rest of the values unchanged:

\setcounter{equation}{0}

\begin{equation}
\mathbf{P}_{\mathbf{S}}(\mathbf{r})= \begin{cases}\rho(\mathbf{r}), & \text { if } \mathbf{r} \in \mathbf{S}, \\ 0, & \text { otherwise }\end{cases}
\label{eqm2}
\end{equation}

Similarly, one can extend this definition to enforce nonnegativity, $\mathbf{P}_{\mathbf{S+}}\rho(\mathbf{r})=\max \left(\mathbf{P}_{\mathbf{S}} \rho(\mathbf{r}), 0\right)$. The modulus projector $\mathbf{P}_\mathbf{M}$ sets the modulus to the retrieved one $m(\mathbf{k})$, and leaving the phase unchanged. 

\begin{equation}
\mathbf{P}_\mathbf{M} \tilde{\rho}(\mathbf{k})=\mathbf{P}_\mathbf{M}|\tilde{\rho}(\mathbf{k})| e^{i \varphi(\mathbf{k})}=m(\mathbf{k}) e^{i \varphi(\mathbf{k})}
\label{eqm3}
\end{equation}

The reflection operators $\mathbf{R}_{\mathbf{S}}$ $=2 \mathbf{P}_{\mathbf{S}}-\mathbf{I}$ and $\mathbf{R}_{\mathbf{M}}=2 \mathbf{P}_{\mathbf{M}}-\mathbf{I}$ apply the same step as the projector but move twice as far, where $\textbf{I}$ means the unit matrix. The phase retrieval algorithm requires a starting point $\rho^0(\mathbf{r})$, generated by assigning a random phase to the retrieved target Fourier magnitude iteratively produce the iterate $\rho^{(n)}(\mathbf{r})$ after $n$ steps. With the notation above, the HIO algorithm can be formulated as

\begin{equation}
\rho^{(n+1)}(\mathbf{r})=\left[\mathbf{P}_\mathbf{S} \mathbf{P}_\mathbf{M}+\mathbf{P}_{\underline{\mathbf{S}}}\left(\mathbf{I}-\beta \mathbf{P}_\mathbf{M}\right)\right] \rho^{(n)}(\mathbf{r})
\label{eqm4}
\end{equation}

Where $\mathbf{P}_{\underline{\mathbf{S}}}=\left(\mathbf{I}-\mathbf{P}_\mathbf{S}\right)$ denotes the complement of the projector $\mathbf{S}$ and the $\beta(0<\beta<1)$ here and below means a negative feedback parameter. Empirically, small $\beta$ produces better stability while $\beta$ is better for escaping local minima. A value of $\beta$ somewhere between 0.5 and 1.0, say 0.8, usually works well. In all experiments and simulations, the $\beta$ is set as 0.8. Similarly, the RAAR algorithm can be written as

\begin{equation}
\rho^{(n+1)}(\mathbf{r})=\left[\frac{1}{2} \beta\left(\mathbf{R}_\mathbf{S} \mathbf{R}_\mathbf{M}+\mathbf{I}\right)+(1-\beta) \mathbf{P}_\mathbf{M}\right] \rho^{(n)}(\mathbf{r}).
\label{eqm5}
\end{equation}

The retrieved target Fourier magnitude (Fig.~\ref{fig: data processing}B) was one-fold zero-padded before entering into the phase retrieval algorithm. For better image performance, a combination of RAAR (60 iterations) and HIO (one iteration) algorithms is employed, with the RAAR escaping local minima and HIO refining the solution. Along with the iterations, two sets of constraints are equipped. The first object constraint was that it had to be real and non-negative. The second object constraint was a shrink-wrap support \cite{marchesini2003x} to refine the support region $\mathbf{S}$ during the iteration. Specifically, the support region $\mathbf{S}$ periodically convolves the iteration with a Gaussian of width $\sigma$ and then applies a threshold $\epsilon$ to smear the support; i.e., the updated support for the two-dimensional case is taken to be 

\begin{equation}
S=\left\{\mathbf{r}:\frac{1}{2 \pi \sigma^2} \int_{\mathbb{R}^2} \rho^{(n)}(\mathbf{s}) e^{-|\mathbf{r}-\mathbf{s}|^2 / 2 \sigma^2} d \mathbf{s} \geq \epsilon\right\}
\label{eqm6}
\end{equation}

The third object constraint used was a sparsity constraint by setting 20 $\%$ of the maximum entries of the original structure. The Fourier space constraint is the magnitude constraint that sets the modulus to the retrieved one. Accompanied by the iteration, the adaptive support region is controlled by an alternative Gaussian convolution and a threshold operation, which act as dilation and erosion. With the iterative number increases, the exported support region will change from a loose one to a tight one, which can significantly reduce the solution space and exclude spurious solutions. It is worth emphasizing that using a non-centrosymmetric initial support estimation, for instance, an equilateral triangle as initial support considerably speeds up convergence and effectively avoids the twin-image problem~\cite{fienup1986phase}.

All experiments were performed with MATLAB 2018b in Windows 10 running on a dual-core chip Intel Core i9-10900K and 128G memory. A single run of the algorithm on a $1024 \times 1024$ pixel image retrieval (composed of 60 (RAAR) +1 (HIO) iterations) on this CPU took approximately 1 s. Notably, all results displayed in experiments and simulations are unthresholded and without color bar alternation.

\section{Imaging performance vs. number of laser emitters.}
\label{methods: different emitters}

\begin{figure}[ht!]
	\centering
	\includegraphics[width=1 \linewidth]{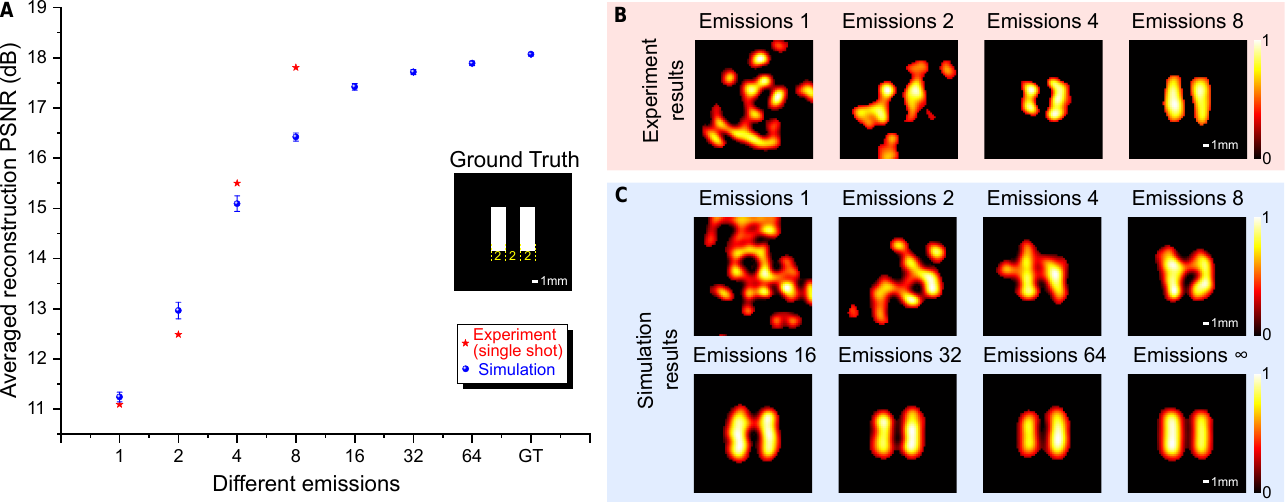}
	\caption{\textbf{Imaging of a double slit with different laser emissions.} The two slits are 2 mm wide and 6 mm long, with a 4 mm center distance between the two slits. (\textbf{A}) The curves of the averaged reconstruction peak SNR versus different emitters ($n$ = 1, 2, 4, 8, 16, 32, 64 and ‘GT’ = Ground truth). In both simulation and experiment, the imaging quality improved as emitters increased. For simulation, in each laser emission group, 50 independent sets of HBT spectrum are sampled (Section \ref{Theory Part III} and Fig.~\ref{fig: Flow chart}A to D) and 50 phase retrieval trials are implemented for each HBT spectrum (Section \ref{methods: retrieval}  and Fig.~\ref{fig: data processing}). The peak SNRs of the 50 $\times$ 50 = 2500 reconstruction results are counted for each emission group, with the error bars indicating the standard deviation of the average reconstruction peak SNR. It should be noted that the results of the experiment are from single-shot measurements. (\textbf{B}) Experimental reconstruction results under different emitters ($n$ = 1, 2, 4, 8). (\textbf{C}) Typical simulation reconstruction results blindly selected from each emission group. Note that the emission ‘$\infty$’ means reconstructed image in a noiseless regime, revealing the theoretical limits of the method. Scale bars: 1 mm ((B), (C)).}
	\label{fig: double slit different emissions}
\addtocounter{Sfigure}{1}
\end{figure}

To evaluate the impact of different laser emitters on image performance, we performed a double slit imaging under different laser emitters with a slit-to-slit separation of approximately 2 mm. The experimental results under $n$ = 1, 2, 4, and 8 emitters as well as the quality performance are shown in Fig.~\ref{fig: double slit different emissions}A and B. Simulations with the same target under $n$ = 1, 2, 4, 8, 16, 32, 64 and ‘$\infty$’ (the ground truth Fourier domain sampling) are carried out and the corresponding results are depicted in Fig.~\ref{fig: double slit different emissions}A and C. In the simulation, we generate 50 sets of data with random initialization for each emission group and perform 50 trials of phase retrieval for each set of data, yielding a total of 50 $\times$ 50 = 2500 reconstructions. For each emission group, the average peak SNRs of the 2500 reconstructions are counted, with the error bars indicating the standard deviation of the mean peak SNR. The thermal features of the light source can be improved by increasing the number of emitters, transforming the double slit from intractable to clearly separated. The experiments and simulations embrace with each other and both indicate that at least four emitters can correctly achieve the remote target. The averaged error bars are fairly modest, especially when the emitters are greater than four. It is worth emphasizing that the experimental results were obtained by single-shot measurements.

\section{Different emissions vs acquisition frames.}

\begin{figure}[ht!]
	\centering
	\includegraphics[width=1 \linewidth]{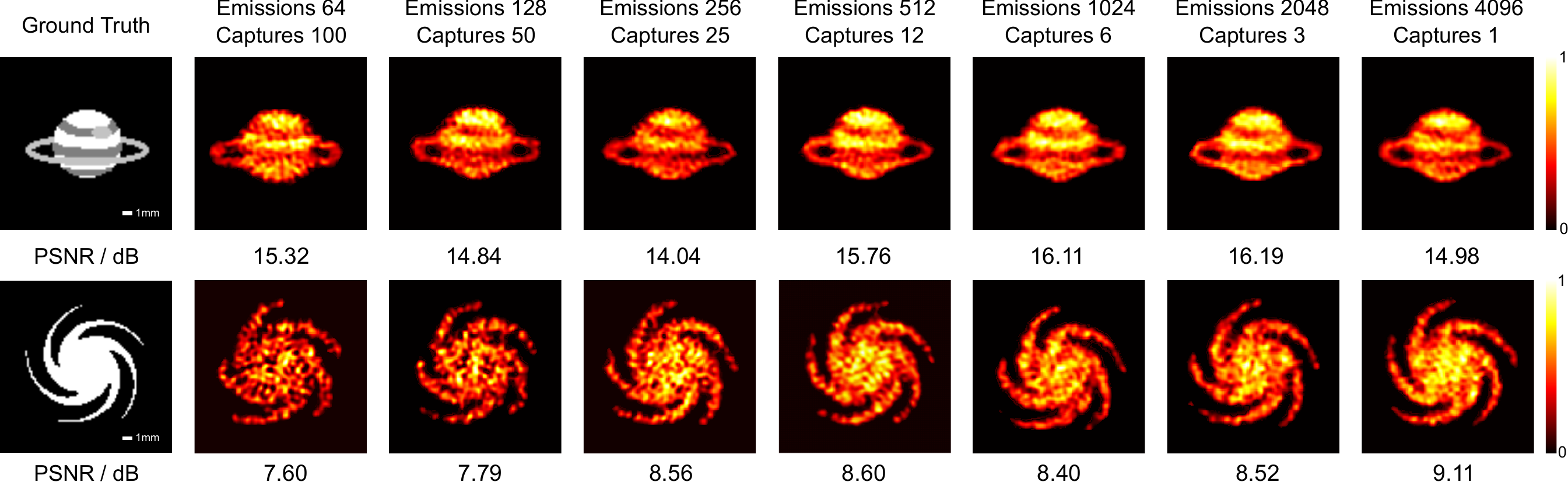}
	\caption{\textbf{Numerical evaluations of the reconstruction with various virtual emissions and average times.} Leftmost column: Ground truth of the simulated targets. Second from left: The impact of increasing laser emissions while decreasing averaging times but maintaining the product of the two essentially constant. Both results show that increasing the number of laser emissions improves image quality slightly more than increasing the average time, indicating that the thermal properties of the light source are more essential to imaging performance (Section \ref{methods: different emitters}). Nonetheless, in the case of restricted emissions, imaging quality can be be improved by increasing the number of captures. The simulation parameter settings are identical to aforementioned except that the virtual baseline shifts from 1 to 127 cm with a fixed interval of $\delta$ = 1 cm. Scale bars: 1 mm (all subgraphs).}
	\label{fig: Different emissions vs acquisition frames}
\addtocounter{Sfigure}{1}
\end{figure}

In the current experimental setup, the incoherent source is modeled by a collection of emitters. Measuring the intensity correlation function $c_{ab}^{(2)}$ with pseudothermal light inevitably leads to speckle-like noise due to interference of the signal with scattered light. By increasing the number of emitters or merging numerous collection frames, speckle-like noise can be reduced while also improving the quality of the returned Fourier magnitude. Two shapes of targets are tested and the reconstruction results with different emissions and acquisition frames from numerical simulation are presented in Fig.~\ref{fig: Different emissions vs acquisition frames}. Both results indicated that increasing the number of laser emissions brings a major enhancement in image quality when compared to accumulating the acquisition frames, demonstrating that the thermal properties of the light source are more influential to the imaging performance. To reduce the speckle-like noise, increasing the number of laser emitters $n$ has a convergence rate of $O(1 / \sqrt{n})$ and this can be further accelerated when the value of the Fourier spectrum approaches to 0 ($O(1 / n)$). However, by accumulating the acquisition frames $k$ has a convergence rate of $1 / \sqrt{k}$ (Section \ref{Theory Part I}). The variation in convergence rate causes disparities in the effectiveness of the two techniques for image quality enhancement. However, if emissions are limited, increasing the number of acquisitions can improve imaging quality.

\section{Comparisons among different phase retrieval algorithms.}

\begin{figure}[ht!]
	\centering
	\includegraphics[width=1 \linewidth]{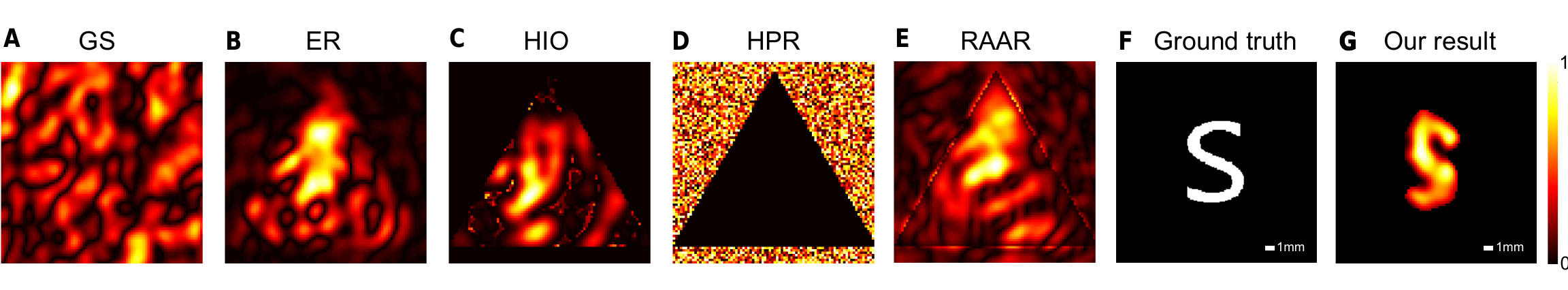}
	\caption{\textbf{Comparisons of the reconstruction performance with different phase retrieval strategies.} (\textbf{A} to \textbf{E}) Reconstructions of the letter ‘S’ with various phase retrieval algorithms, including the algorithm Gerchberg-Saxton (GS), Error Reduction (ER), Hybrid Input Output (HIO), Hybrid Projection Reflection (HPR), Relaxed Averaged Alternating Reflectors (RAAR). Note that the iterative process of (A) to (E) does not employ the sparse constraint and the adaptive shrink-wrap support (see section of Image Reconstruction and Fig.~\ref{fig: data processing}D). (\textbf{F}) The ground truth target. (\textbf{G}) The results obtained from the proposed method. Scale bars: 1 mm ((F), (G)).}
	\label{fig: Different PR}
\addtocounter{Sfigure}{1}
\end{figure}

To demonstrate the superiority of the proposed phase retrieval scheme, we conducted an experimental comparison by imaging the letter ‘S’ (Fig.~\ref{fig: Different PR}F and Fig.~\ref{fig: exp results}D). The letter ‘S' was chosen because curve shapes contain more complex features in Fourier space. However, the retrieved target Fourier magnitude (Fig.~\ref{fig: exp results}C) is still noise-stained and has fainter high-frequency information (Fig.~\ref{fig: exp results}C). The requirements for applying the Gerchberg-Saxton (GS) \cite{gerchberg1972practical} iterative phase retrieval algorithm and its derivations such as Error Reduction (ER) \cite{levi1984image} and Hybrid Input Output (HIO), are not satisfied and the resulting images (Fig.~\ref{fig: Different PR}A to C) are of poor quality. More advanced alternating projection algorithms such as Hybrid Projection Reflection (HPR) \cite{bauschke2003hybrid} (Fig.~\ref{fig: Different PR}D) and Relaxed Averaged Alternating Reflectors (RAAR) (Fig.~\ref{fig: Different PR}E), are also unable to recover the target. In stark contrast, the image reconstructed from the proposed method (Fig.~\ref{fig: Different PR}G) significantly outperforms the others, with a diffraction-limited resolution of 1.5 mm (limited by the virtual aperture of 87 cm). High-fidelity reconstructions are attributed to the majority voting and reduced support constraint as well as the cascade scheme. For the algorithms compared, the number of iterations was set to 60, the initial support was used as a non-centrosymmetric triangle, and the negative feedback parameter $\beta$ for the single-cycle HIO, HPR, and RAAR algorithms was fixed at 0.8.

\section{Complex targets imaging via active intensity interferometry.}

\begin{figure}[ht!]
	\centering
	\includegraphics[width=1 \linewidth]{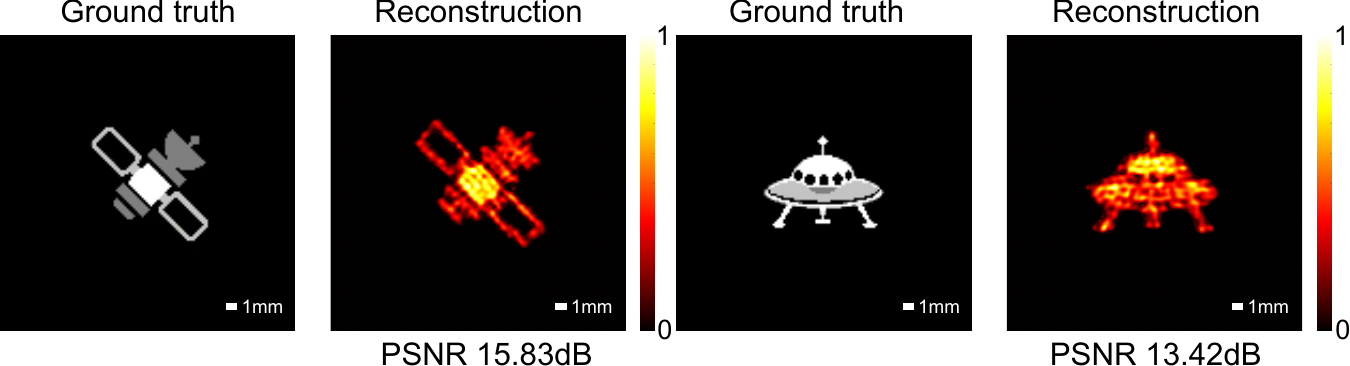}
	\caption{\textbf{Numerical evaluations of the imaging capability of complex targets with varying levels of reflectivity.} (\textbf{A}), (\textbf{C}), Ground truth of the simulated targets. (\textbf{B}), (\textbf{D}), Reconstruction results from single-shot data acquisition via the active intensity interferometer. In both situations, the reconstructed irradiance is nearly identical to the true value. The virtual baseline shifts from 1 to 247 cm with a fixed interval of $\delta$ = 1 cm. The remaining parameters are the same as in the experiment. With sufficient Fourier space sampling, our technique has the capacity to observe complex targets as well as recover their surfaces with varying reflectivity levels. Scale bars: 1 mm ((A) to (D)).}
	\label{fig: complex targets}
\addtocounter{Sfigure}{1}
\end{figure}

In this proof of concept, we have demonstrated two-dimensional superresolution imaging over kilometers range with millimeters precision via the proposed active intensity interferometry. Higher-fidelity complex target reconstruction is anticipated by improving the sampling rate on the interferometric plane and multiplying the number of baselines/telescopes. To demonstrate this, complex targets (the targets ‘satellite’ and ‘UFO’) with varying reflectivity levels imaged under a 247 cm baseline with a 1 cm sampling interval were tested, and the corresponding results are depicted in Fig.~\ref{fig: complex targets}. Since the intensity interferometry connects telescopes only electronically, it is practically insensitive to atmospheric turbulence and optical imperfections, permitting observations over very long baselines and through large airmasses, as well as at short optical wavelengths. With kilometer baselines, sufficient sampling frequencies, and advanced algorithms, tantalizing results are expected to open the window toward new scientific results.

\section{Theory Part I. Theory of multiple laser emitters with thermal nature.}
\label{Theory Part I}

We model our imaging system based on statistical optics. As shown in Fig.~\ref{fig: forward model} in the main text, suppose our optical system consists of $n$ laser emitters with angular frequency $\omega$ numbered $1\sim n$ in the source plane, a target that is a scattering sample of a particular structure in the object plane, and two movable detectors $D_a$ and $D_b$ in the detection plane. Detectors $D_a$ and $D_b$ always use the same polarization measurement in the experiment, then we can denote the electric fields they measure on this polarization component at time $t$ as $E_a(t)$ and $E_b(t)$, respectively. For simplicity, we only consider the case of far-field illumination and detection, that is, the distance between the target and the emitters or detectors is much greater than the Fraunhofer distance $2D^2/\lambda$, where $\lambda$ is the laser wavelength and $D$ is the maximum value among laser beam waists, detector apertures and target size. 

If $D$ is much smaller than the atmospheric coherence length, we can approximate each emitter's illumination light and its scattered light scattered by the target as an intensity- and phase-modulated plane wave. To describe these plane waves, we denote $\textbf{k}_i$ as the wave vector from the $i$-th emitter to the target, and denote $\textbf{k}_a$ and $\textbf{k}_b$ as the wave vectors from the target to the detector $D_a$ and $D_b$, respectively. Combining the above approximations and further ignoring all CW laser delays due to the time-of-flight which is much smaller than the atmospheric coherence time, we can approximately express $E_a(t)$ and $E_b(t)$ as
\begin{equation}
E_{a,b}(t)=\sum_{i=1}^{n}\sqrt{I_i(t)} \mathrm{e}^{\mathrm{i}(\phi_{i}(t)-\omega t)}T(\textbf{k}_i,\textbf{k}_{a,b})T_{a,b}(t).
\end{equation}
Here, $I_i(t)$ and $\phi_{i}(t)$ represent the intensity and phase of the plane wavefront formed at the target at time $t$ by the illumination light from the $i$th emitter, $T(\textbf{k}_{in},\textbf{k}_{out})$ is the target's transmission matrix between the $\textbf{k}_{in}$ input field and $\textbf{k}_{out}$ output field and $T_a(t), T_b(t)$ are the field propagation coefficients from the target to the detectors $D_a$ and $D_b$ at time $t$, respectively. All the $I_i(t),\phi_{i}(t)$ and $T_a(t), T_b(t)$ are treated as time-varying random variables which incorporate the whole process of atmospheric disturbance. 

Considering that it is difficult to make a complete statistical process modeling of atmospheric disturbance, we approximate the impact of atmospheric disturbance as the following two simple statistical assumptions. The first assumption is that all the $I_i(t),\phi_{i}(t)$ and $T_a(t), T_b(t)$ are independent of each other. Note that this assumption clearly does not hold when the distances between different emitters or detectors are smaller than the atmospheric coherence length, so we need to avoid this in the experiment. The second assumption is that all the $\phi_{i}(t)$ are Gaussian random processes. Let the notation $\langle ...\rangle_t$ denote the time average over a sufficiently long time. Define $\overline{\phi_i}\equiv\langle\phi_{i}(t)\rangle_t$ and $\Delta\phi_{i}(t)\equiv\phi_{i}(t)-\overline{\phi_i}$, using the properties of Gaussian distribution it's easy to prove that
\begin{equation}
\label{gaussian}
\langle\mathrm{e}^{\mathrm{i}(\phi_{i}(t)-\phi_{j}(t))}\rangle_t=\mathrm{e}^{\mathrm{i}(\overline{\phi_i}-\overline{\phi_j})-\frac{1}{2}\langle(\Delta\phi_{i}(t)-\Delta\phi_{j}(t))^2\rangle_t}.
\end{equation}
In this way, if $\langle(\Delta\phi_{i}(t)-\Delta\phi_{j}(t))^2\rangle_t\gg 1$ when $i\neq j$, we can further have an approximation of
\begin{equation}
\label{phase}
\langle\mathrm{e}^{\mathrm{i}(\phi_{i}(t)-\phi_{j}(t))}\rangle_t=\delta_{ij}\equiv\left\{
\begin{array}{rcl}
1 && {i=j}\\
0 && {i\neq j}
\end{array}.
\right.
\end{equation}
This approximation is valid when the distance between emitters is much greater than the atmospheric coherence length. In fact, the $\langle(\Delta\phi_{i}(t)-\Delta\phi_{j}(t))^2\rangle_t\approx 1$ condition can be used as a definition of atmospheric coherence length. For example, the Kolmogorov turbulence model\cite{fried1966optical,ridley2011measurements} expected that
\begin{equation}
\label{Kolmogorov}
\langle(\Delta\phi_{i}(t)-\Delta\phi_{j}(t))^2\rangle_t= 6.88(\frac{r_{ij}}{r_0})^{\frac{5}{3}},
\end{equation}
where $r_{ij}$ is the distance between the $i$-th emitter and the $j$-th emitter, $r_0$ is the Fried parameter, which can be used as a representation of the atmospheric coherence length.

Define $I_a(t)\equiv|E_a(t)|^2$ and $I_b(t)\equiv|E_b(t)|^2$ as the intensities measured by detectors $D_a$ and $D_b$ at time $t$, respectively, after using the above assumptions, the time-averaged intensity can be expressed as
\begin{equation}
\label{tai}
\langle I_{a,b}(t)\rangle_t=\langle|E_{a,b}(t)|^2\rangle_t=\eta_{a,b}\sum_{i=1}^{n}\overline{I_i}|T(\textbf{k}_i,\textbf{k}_{a,b})|^2,
\end{equation}
where $\eta_{a,b}\equiv\langle|T_{a,b}(t)|^2\rangle_t$ and $\overline{I_i}\equiv\langle I_i(t)\rangle_t$. Define $c_i^{(2)}\equiv\langle I_i^2(t)\rangle_t/\overline{I_i}^2-1$, we further have the intensity correlation function
\begin{equation}
\begin{aligned}
\label{icf}
\langle\Delta I_a(t)\Delta I_b(t)\rangle_t&\equiv\langle I_a(t)I_b(t)\rangle_t-\langle I_a(t)\rangle_t\langle I_b(t)\rangle_t\\
&=\eta_a\eta_b[\sum_{i\neq j}\overline{I_i}\,\overline{I_j}T(\textbf{k}_i,\textbf{k}_a)T(\textbf{k}_j,\textbf{k}_b)T^*(\textbf{k}_i,\textbf{k}_b)T^*(\textbf{k}_j,\textbf{k}_a)\\
&+\sum_{i=1}^{n}c_i^{(2)}\overline{I_i}^2|T(\textbf{k}_i,\textbf{k}_a)|^2|T(\textbf{k}_i,\textbf{k}_b)|^2],
\end{aligned}
\end{equation}

It can be seen that for a particular target, both the time-averaged intensity and the intensity correlation function have a particular value determined by the transmission matrix $T(\textbf{k}_{in},\textbf{k}_{out})$ of the scattering sample. Now we consider an ensemble of scattering samples with the same geometric surface but different internal microstructure. The fully developed speckle model proposed by Goodman et al. predicts that if the sample surface is sufficiently rough compared to the wavelength of light, for each $(\textbf{k}_{in},\textbf{k}_{out})$ pair, the transmission coefficient $T(\textbf{k}_{in},\textbf{k}_{out})$ behaves as a circular complex Gaussian random variable under different ensembles \cite{goodman2007speckle}. 

We next demonstrate that this statistical property of the transmission matrix is the key factor causing speckle-like noise in the experimental data. Let the notation $\langle...\rangle_e$ denote the ensemble average, and the notation $\langle...\rangle_{t,e}\equiv\langle\langle...\rangle_t\rangle_e$ denote the ensemble average of a time average quantity. Using the moment theorem \cite{reed1962moment} for complex Gaussian random variables, we have the ensemble-averaged intensity correlation function
\begin{equation}
\begin{aligned}
\label{eaicf}
&\langle\Delta I_a(t)\Delta I_b(t)\rangle_{t,e}
=\eta_a\eta_b[\sum_{i\neq j}\overline{I_i}\,\overline{I_j}\langle T(\textbf{k}_i,\textbf{k}_a)T^*(\textbf{k}_i,\textbf{k}_b)\rangle_e\langle T(\textbf{k}_j,\textbf{k}_b)T^*(\textbf{k}_j,\textbf{k}_a)\rangle_e\\
&+\sum_{i\neq j}\overline{I_i}\,\overline{I_j}\langle T(\textbf{k}_i,\textbf{k}_a)T^*(\textbf{k}_j,\textbf{k}_a)\rangle_e \langle T(\textbf{k}_j,\textbf{k}_b)T^*(\textbf{k}_i,\textbf{k}_b)\rangle_e\\
&+\sum_{i=1}^{n}c_i^{(2)}\overline{I_i}^2(\langle|T(\textbf{k}_i,\textbf{k}_a)|^2\rangle_e \langle|T(\textbf{k}_i,\textbf{k}_b)|^2\rangle_e+|\langle T(\textbf{k}_i,\textbf{k}_a)T^*(\textbf{k}_i,\textbf{k}_b)\rangle_e|^2)].
\end{aligned}
\end{equation}

To further obtain an explicit expression of the above formula, we need to study the coherence function of the form $\langle T(\textbf{k}_{in1},\textbf{k}_{out1})T^*(\textbf{k}_{in2},\textbf{k}_{out2})\rangle_e$, where $\textbf{k}_{in1}=\textbf{k}_{in2}$ or $\textbf{k}_{out1}=\textbf{k}_{out2}$. For the first case, the classical coherence theory based on the van Cittert-Zernike theorem gives \cite{goodman2015statistical}
\begin{equation}
\langle T(\textbf{k}_{in},\textbf{k}_{out1})T^*(\textbf{k}_{in},\textbf{k}_{out2})\rangle_e
\propto \int\rho(\textbf{r})\mathrm{e}^{-\mathrm{i}(\textbf{k}_{out1}-\textbf{k}_{out2}).\textbf{r}}\mathrm{d}\textbf{r},
\end{equation}
which is the Fourier transform of the intensity distribution on the output surface of the target. So let
\begin{equation}
\label{coh1}
\langle T(\textbf{k}_{in},\textbf{k}_{out1})T^*(\textbf{k}_{in},\textbf{k}_{out2})\rangle_e=\eta f(\textbf{k}_{out1}-\textbf{k}_{out2}),
\end{equation}
where $\eta$ is a real constant and $f$ is the normalized Fourier function, that is,
\begin{equation}
f(\Delta\textbf{k})=\frac{\int\rho(\textbf{r})\mathrm{e}^{-\mathrm{i}\Delta\textbf{k}.\textbf{r}}\mathrm{d}\textbf{r}}{\int\rho(\textbf{r})\mathrm{d}\textbf{r}}.
\end{equation}
For the second case, we assume that
\begin{equation}
\label{coh2}
\langle T(\textbf{k}_{in1},\textbf{k}_{out})T^*(\textbf{k}_{in2},\textbf{k}_{out})\rangle_e=\eta h(\textbf{k}_{in1}-\textbf{k}_{in2}),
\end{equation}
where $h$ is a function that satisfies $h(0)=f(0)=1.$ For surface scattering, classical coherence theory simply predicts $h=f$, but for volume scattering, the physical mechanism here is much more complicated. Some modern studies on speckle have found that the function $h$ is related to the optical memory effect. Both theory \cite{feng1988correlations} and experiment \cite{freund1988memory} show that $h$ will decay to 0 when $|\textbf{k}_{i}-\textbf{k}_{j}|$ is large, and we will not discuss the relevant details here. Substituting Eq. (\ref{coh1}) and Eq. (\ref{coh2}) into Eq. (\ref{tai}) and Eq. (\ref{eaicf}), we obtain
\begin{equation}
\label{etai}
\langle I_{a,b}(t)\rangle_{t,e}=\eta_{a,b}\,\eta\sum_{i=1}^{n}\overline{I_i}
\end{equation}
and
\begin{equation}
\label{etaicf}
\begin{aligned}
\langle\Delta I_a(t)\Delta I_b(t)\rangle_{t,e}&=\eta_a\eta_b\,\eta^2[\sum_{i=1}^{n}c_i^{(2)}\overline{I_i}^2+\sum_{i\neq j}\overline{I_i}\,\overline{I_j}|h(\textbf{k}_i-\textbf{k}_j)|^2\\
&+(\sum_{i=1}^{n}c_i^{(2)}\overline{I_i}^2+\sum_{i\neq j}\overline{I_i}\,\overline{I_j})|f(\textbf{k}_a-\textbf{k}_b)|^2],
\end{aligned}
\end{equation}
respectively. In our experiment, the quantity we measure is the normalized intensity correlation function $c_{ab}^{(2)}$, which is defined as
\begin{equation}
c_{ab}^{(2)}\equiv\frac{\langle\Delta I_a(t)\Delta I_b(t)\rangle_t}{\langle I_a(t)\rangle_{t,e}\langle I_b(t)\rangle_{t,e}}.
\end{equation}
According to Eq. (\ref{etai}) and Eq. (\ref{etaicf}), the ensemble average of $c_{ab}^{(2)}$ is
\begin{equation}
\langle c_{ab}^{(2)}\rangle_e=c_0+c_1|f(\textbf{k}_a-\textbf{k}_b)|^2,
\end{equation}
where
\begin{equation}
c_0=\frac{\sum_{i=1}^{n}c_i^{(2)}\overline{I_i}^2+\sum_{i\neq j}\overline{I_i}\,\overline{I_j}|h(\textbf{k}_i-\textbf{k}_j)|^2}{(\sum_{i=1}^{n}\overline{I_i})^2},
\end{equation}
and
\begin{equation}
c_1=\frac{\sum_{i=1}^{n}c_i^{(2)}\overline{I_i}^2+\sum_{i\neq j}\overline{I_i}\,\overline{I_j}}{(\sum_{i=1}^{n}\overline{I_i})^2}.
\end{equation}

To quantify the level of speckle-like noise, we further calculate $\sigma(c_{ab}^{(2)})$, which is the standard deviation of $c_{ab}^{(2)}$, defined as

\begin{equation}
\sigma(c_{ab}^{(2)})\equiv\sqrt{\langle(c_{ab}^{(2)}-\langle c_{ab}^{(2)}\rangle_e)^2\rangle_e} .
\end{equation}
If the optical memory effect of the scattering sample is neglected, we can show that
\begin{equation}
\sigma(c_{ab}^{(2)})=\sqrt{d_0+d_1|f(\textbf{k}_a-\textbf{k}_b)|^2+d_2|f(\textbf{k}_a-\textbf{k}_b)|^4},
\end{equation}
where
\begin{equation}
d_0=\frac{\sum_{i=1}^{n}3(c_i^{(2)}\overline{I_i}^2)^2+\sum_{i\neq j}(\overline{I_i}\,\overline{I_j})^2}{(\sum_{i=1}^{n}\overline{I_i})^4},
\end{equation}
\begin{equation}
d_1=\frac{\sum_{i=1}^{n}14(c_i^{(2)}\overline{I_i}^2)^2+\sum_{i\neq j}(12c_i^{(2)}\overline{I_i}^3\overline{I_j}+2\overline{I_i}^2\overline{I_j}^2)+\sum_{i\neq j,j\neq k,k\neq i}2\overline{I_i}^2\overline{I_j}\,\overline{I_k}}{(\sum_{i=1}^{n}\overline{I_i})^4},
\end{equation}
and 
\begin{equation}
d_2=\frac{\sum_{i=1}^{n}3(c_i^{(2)}\overline{I_i}^2)^2+\sum_{i\neq j}(4c_i^{(2)}\overline{I_i}^3\overline{I_j}+3\overline{I_i}^2\overline{I_j}^2)+\sum_{i\neq j,j\neq k,k\neq i}2\overline{I_i}^2\overline{I_j}\,\overline{I_k}}{(\sum_{i=1}^{n}\overline{I_i})^4}.
\end{equation}
For the derivation details of the above formula, see Section \ref{Theory Part II}.

To better understand these results from a physical perspective, we consider two simplified cases. The first case is to use only one laser emitter for illumination, that is, $n=1$, then the ensemble average and the standard deviation of $c_{ab}^{(2)}$ can be simplified as
\begin{equation}
\langle c_{ab}^{(2)}\rangle_e=c_1^{(2)}(1+|f(\textbf{k}_a-\textbf{k}_b)|^2)
\end{equation}
and 
\begin{equation}
\begin{aligned}
\sigma(c_{ab}^{(2)})=c_1^{(2)}(3+14|f(\textbf{k}_a-\textbf{k}_b)|^2+3|f(\textbf{k}_a-\textbf{k}_b)|^4)^{\frac{1}{2}},
\end{aligned}
\end{equation}
respectively. In the ideal situation without atmospheric disturbance, $c_1^{(2)}=0$, then $c_{ab}^{(2)}=\langle c_{ab}^{(2)}\rangle_e=\sigma(c_{ab}^{(2)})=0$ can be further obtained, which is in line with the expectation that no intensity interference signal can be detected under ideal coherent light illumination. When atmospheric disturbance can cause an autocorrelation coefficient $c_1^{(2)}$ that cannot be ignored, an intensity interference signal proportional to $c_1^{(2)}$ on average can actually be measured. However, since $\sigma(c_{ab}^{(2)})$ is also proportional to $c_1^{(2)}$, the level of speckle-like noise is almost the same as the signal (easy to prove that $\sqrt{3}<\sigma(c_{ab}^{(2)})/\langle c_{ab}^{(2)}\rangle_e<\sqrt{5}$), resulting in a very low SNR.

The second case is when there is more than one emitter but all emitters have identical properties and the optical memory effect of the scattering sample is neglected. We assume $\overline{I_1}=\overline{I_2}=...=\overline{I_n}$ and $c_1^{(2)}=c_2^{(2)}=...=c_n^{(2)}=c$ is an autocorrelation coefficient, then we have the expressions
\begin{equation}
\langle c_{ab}^{(2)}\rangle_e=\frac{c}{n}+\frac{n-1+c}{n}|f(\textbf{k}_a-\textbf{k}_b)|^2
\end{equation}
and 
\begin{equation}
\begin{aligned}
\sigma(c_{ab}^{(2)})&=[\frac{n-1+3c^2}{n^3}+\frac{2n^2+(12c-4)n+14c^2-12c+2}{n^3}|f(\textbf{k}_a-\textbf{k}_b)|^2\\
&+\frac{2n^2+(4c-3)n+3c^2-4c+1}{n^3}|f(\textbf{k}_a-\textbf{k}_b)|^4]^{\frac{1}{2}},
\end{aligned}
\end{equation}
respectively. We can see that as the number of laser emitters $n$ increases, $\langle c_{ab}^{(2)}\rangle_e$ will approach $|f(\textbf{k}_a-\textbf{k}_b)|^2$, and $\sigma(c_{ab}^{(2)})$ will decrease to 0 at a rate of $O(1/\sqrt{n})$, that is, the speckle-like noise will gradually disappear. When $n$ approaches infinity, our theory and the classical HBT theory based on ideal thermal light sources will give consistent predictions.

\section{Theory Part II. Detailed derivation of the $\sigma(c_{ab}^{(2)})$.}
\label{Theory Part II}

To simplify the calculation of $\sigma(c_{ab}^{(2)})$, we first introduce the notation $X_1\sim X_n$, which are defined as
\begin{equation}
X_i\equiv\overline{I_i} T(\textbf{k}_i,\textbf{k}_a)T^*(\textbf{k}_i,\textbf{k}_b)\,\ (\forall i \in \{1,2,...,n\}).
\end{equation}
Then $c_{ab}^{(2)}$ can be rewritten as
\begin{equation}
\label{cabrw}
c_{ab}^{(2)}=\frac{\sum_{i\neq j}X_i X_j^*+\sum_{i=1}^{n}c_i^{(2)}|X_i|^2}{\eta^2(\sum_{i=1}^{n}\overline{I_i})^2}.
\end{equation}

We also introduce notations $D$ and $Cov$ to represent variance and covariance under ensemble statistics, that is, if $X$ and $Y$ are random variables, we define
\begin{equation}
\begin{aligned}
D(X)&\equiv\langle X^2\rangle_e-\langle X\rangle_e^2,\\
Cov(X,Y)&\equiv\langle XY\rangle_e-\langle X\rangle_e\langle Y\rangle_e.
\end{aligned}
\end{equation}
In this way, $\sigma(c_{ab}^{(2)})$ can be expressed as
\begin{equation}
\label{scabexpand}
\begin{aligned}
\sigma(c_{ab}^{(2)})&=\frac{\sqrt{D(\sum_{i\neq j}X_i X_j^*+\sum_{i=1}^{n}c_i^{(2)}|X_i|^2)}}{\eta^2(\sum_{i=1}^{n}\overline{I_i})^2}\\
&=\frac{\sqrt{D(\sum_{i\neq j}X_i X_j^*)+2Cov(\sum_{i\neq j}X_i X_j^*,\sum_{i=1}^{n}c_i^{(2)}|X_i|^2)+D(\sum_{i=1}^{n}c_i^{(2)}|X_i|^2)}}{\eta^2(\sum_{i=1}^{n}\overline{I_i})^2}
\end{aligned}
\end{equation}

Using the momentum theorem \cite{reed1962moment}, we can simply prove that for any $i=1\sim n$, $X_i$ has the following properties:
\begin{equation}
\begin{aligned}
\langle X_i\rangle_e&=\eta\overline{I_i}f(\textbf{k}_a-\textbf{k}_b),\\
\langle X_i^2\rangle_e&=2\eta^2\overline{I_i}^2f^2(\textbf{k}_a-\textbf{k}_b),\\
\langle |X_i|^2\rangle_e&=\eta^2\overline{I_i}^2(1+|f(\textbf{k}_a-\textbf{k}_b)|^2),\\
\langle X_i|X_i|^2\rangle_e&=2\eta^3\overline{I_i}^3f(\textbf{k}_a-\textbf{k}_b)(2+|f(\textbf{k}_a-\textbf{k}_b)|^2),\\
\langle |X_i|^4\rangle_e&=4\eta^4\overline{I_i}^4(1+4|f(\textbf{k}_a-\textbf{k}_b)|^2+|f(\textbf{k}_a-\textbf{k}_b)|^4).\\
\end{aligned}
\end{equation}
On the other hand, if the optical memory effect of the scattering sample is neglected, it means
\begin{equation}
\langle T(\textbf{k}_{in1},\textbf{k}_{out1})T^*(\textbf{k}_{in2},\textbf{k}_{out2})\rangle_e=0\,\ (\forall \textbf{k}_{in1}\neq \textbf{k}_{in2}).
\end{equation}
It can be immediately proved that $X_1\sim X_n$ are independent of each other, that is, any multivariate moment can always be decomposed into the form of the product of univariate moments, and its mathematical expression is

\begin{equation}
\langle\prod_{i=1}^{n}X_i^{\alpha_i}X_i^{*\beta_i}\rangle_e=\prod_{i=1}^{n}\langle X_i^{\alpha_i}X_i^{*\beta_i}\rangle_e,
\end{equation}
where $\alpha_1\sim\alpha_n$ and $\beta_1\sim\beta_n$ are any non-negative integers.

Using the above conclusions, we can do the following calculations:
\begin{equation}
\begin{aligned}
D(\sum_{i\neq j}X_i X_j^*)&=\sum_{i\neq j}(D(X_i X_j^*)+Cov(X_i X_j^*,X_j X_i^*))\\
+&\sum_{i\neq j,j\neq k,k\neq i}Cov(X_i X_j^*,(X_i+X_j)X_k^*+X_k(X_i^*+X_j^*))\\
&=\sum_{i\neq j}\eta^4\overline{I_i}^2\overline{I_j}^2(1+2|f(\textbf{k}_a-\textbf{k}_b)|^2+3|f(\textbf{k}_a-\textbf{k}_b)|^4)\\
+&\sum_{i\neq j,j\neq k,k\neq i}\eta^4\overline{I_i}\,\overline{I_j}(\overline{I_i}+\overline{I_j})\overline{I_k}(|f(\textbf{k}_a-\textbf{k}_b)|^2+|f(\textbf{k}_a-\textbf{k}_b)|^4),\\
Cov(\sum_{i\neq j}X_i X_j^*,\sum_{i=1}^{n}c_i^{(2)}|X_i|^2)&=\sum_{i\neq j}Cov(X_i X_j^*,c_i^{(2)}|X_i|^2+c_j^{(2)}|X_j|^2)\\
&=\sum_{i\neq j}\eta^4\overline{I_i}\,\overline{I_j}(c_i^{(2)}\overline{I_i}^2+c_j^{(2)}\overline{I_j}^2)(3|f(\textbf{k}_a-\textbf{k}_b)|^2+|f(\textbf{k}_a-\textbf{k}_b)|^4),\\
D(\sum_{i=1}^{n}c_i^{(2)}|X_i|^2)&=\sum_{i=1}^{n}D(c_i^{(2)}|X_i|^2)\\
&=\sum_{i=1}^{n}\eta^4(c_i^{(2)}\overline{I_i}^2)^2(3+14|f(\textbf{k}_a-\textbf{k}_b)|^2+3|f(\textbf{k}_a-\textbf{k}_b)|^4)
\end{aligned}
\end{equation}
Substituting them into Eq. (\ref{scabexpand}) yields the expression of $\sigma(c_{ab}^{(2)})$.

\section{Theory Part III. Data acquisition simulation of the active intensity interferometry.}
\label{Theory Part III}

This chapter will discuss how to construct an algorithm for simulating $c_{ab}^{(2)}$ sampling based on the aformentioned statistical optical model. For simplicity, we take all physical quantities to be dimensionless quantities, and only simulate the case where all laser emitters are symmetrically equivalent, that is, $\eta=1$ (dimensionless), $\overline{I_1}=\overline{I_2}=...=\overline{I_n}=1$ (dimensionless) and $c_1^{(2)}=c_2^{(2)}=...=c_n^{(2)}=c$ (autocorrelation coefficent). Substituting them into Eq. (\ref{cabrw}), we have

\begin{equation}
c_{ab}^{(2)}=\frac{|\sum_{i=1}^{n}X_i|^2+(c-1)\sum_{i=1}^{n}|X_i|^2}{n^2},
\end{equation}

where $X_i=T(\textbf{k}_i,\textbf{k}_a)T^*(\textbf{k}_i,\textbf{k}_b)$ for any $i=1\sim n$ and all the $T(\textbf{k}_i,\textbf{k}_{a,b})\sim\mathcal{CN}(0,1)$ are complex standard normal random variables. We futher ignore the optical memory effect of the scattering sample, then according to the discussion in the previous chapter, all the $X_i$ are independent random variables. In this case, the algorithm only needs to independently sample the random variable $X$ a total of $n$ times to obtain $X_1\sim X_n$, and then generate a simulating sample of the $c_{ab}^{(2)}$.

Now we discuss how to construct an algorithm to sample $X$. We rewrite $X$ as $X=c_1c_2^*$, where $c_1\sim\mathcal{CN}(0,1)$ and $c_2\sim\mathcal{CN}(0,1)$ satisfy $\langle c_1 c_2^*\rangle_e=f(\textbf{k}_a-\textbf{k}_b)$. Using the linear combination of circular complex Gaussian random variables still satisfying the properties of circular complex Gaussian distribution, we can sample $c_1$ and $c_2$ in the following way:

\begin{equation}
\left\{
\begin{array}{rcl}
\begin{aligned}
c_1&=\sqrt{1-|f(\textbf{k}_a-\textbf{k}_b)|^2}z_1+f(\textbf{k}_a-\textbf{k}_b)z_2 \\
c_2&=z_2 
\end{aligned}
\end{array}.
\right.
\end{equation}

Here, $z_1\sim\mathcal{CN}(0,1)$ and $z_2\sim\mathcal{CN}(0,1)$ are independent of each other. On a computer that can only executes the programs with real numbers, the program can independently sample $x\sim\mathcal{N}(0,1/2)$ and $y\sim\mathcal{N}(0,1/2)$ and let $z=x+\mathrm{i}y$ be used as a sampling of a complex standard normal random variable. Combining all of the above, we finally construct an algorithm for sampling  $c_{ab}^{(2)}$.

\section{Estimation and Optimization of SNR in Coincidence Counting Measurement}
\label{Estimation and Optimization of SNR in Coincidence Counting Measurement}
In addition to the speckle-like noise discussed in the previous sections, for a practical intensity interferometer, the randomness of the coincidence counting measurement will also introduce noise into the intensity interference signal. For classical stellar intensity interferometers, the SNR has been shown to be estimated as\cite{malvimat2014intensity}
\begin{equation}
\label{snr1}
\text{SNR}\sim r\Delta\tau\sqrt{\frac{T}{\Delta t}},
\end{equation}
where $r$ is the count rate of the photon detectors (number of photons per unit time), $\Delta\tau$ is the coherence time of the light source, $\Delta t$ is the time bin for coincidence counting (or the time resolution of the photon detectors) and $T$ is the integration time. Eq. (\ref{snr1}) uses two important assumptions in the derivation process, one is that the time resolution of the photon detectors are much larger than the coherence time of the light source, that is, $\Delta t\gg\Delta\tau$, and the other is the Poisson noise approximation when the photon count rate is low, that is, $r\Delta\tau\ll1$. These two assumptions are generally valid for stellar intensity interferometers whose observation targets are weak and broad-spectrum light sources.

However, for active intensity interferometry, both assumptions may not hold because of the use of pseudothermal illumination. Let us first consider only the case where the first assumption does not hold. When $\Delta t\gg\Delta\tau$, the visibility of the intensity interference signal is of $\Delta\tau/\Delta t$ magnitude, but when $\Delta t\ll\Delta\tau$, the maximum visility will be exactly 1. So reconsidering the derivation process of Eq. (\ref{snr1}) in the origin article\cite{malvimat2014intensity}, when $\Delta t\ll\Delta\tau$, the SNR need to be estimated as

\begin{equation}
\label{snr2}
\text{SNR}\sim r\sqrt{T\Delta t}.
\end{equation}

Combining the above two different boundary conditions, it can be found that the SNR is maximized when $\Delta t$ and $\Delta \tau$ are of the same order of magnitude.

In addition, when the count rate increases, the SNR will also increase proportionally, but it cannot be infinitely improved. When $r^2\Delta\tau\Delta t\gg 1$, on average, $r^2\Delta\tau\Delta t$ coincidence events will occur every $\Delta\tau$ time, and they are almost all correlated. That is to say, every $r^2\Delta\tau\Delta t$ consecutive coincidence events can only be approximately combined into one ``effective'' coincidence event. Therefore, the SNR given by Eq. (\ref{snr1}) or Eq. (\ref{snr2}) is overestimated by a factor of $r\sqrt{\Delta\tau\Delta t}$ considering that the SNR is proportional to the square root of the total coincidence rate. Ultimately, the theoretical maximum value of SNR can be expressed as

\begin{equation}
\label{snr3}
\text{max(SNR)}\sim\left\{
\begin{array}{rcl}
r\sqrt{T\Delta\tau} && {\Delta t\approx\Delta\tau\text{ when }r\Delta\tau\ll1}\\
\sqrt{\frac{T}{\Delta\tau}}. && {\frac{1}{r^2\Delta\tau}\ll\Delta t\ll\Delta\tau\text{ when }r\Delta\tau\gg1}
\end{array}.
\right.
\end{equation}

\begin{figure}[ht!]
	\centering
	\includegraphics[width=0.8 \linewidth]{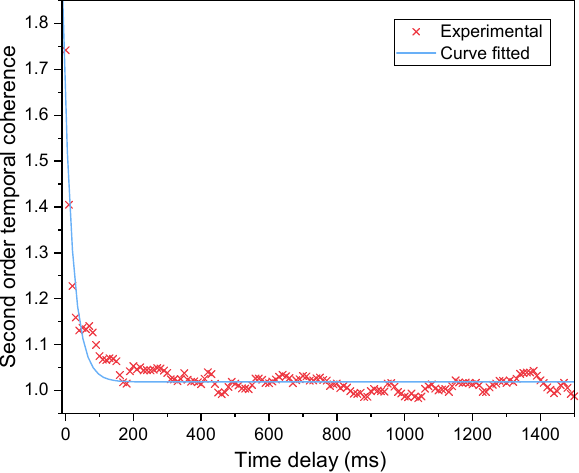}
	\caption{\textbf{Estimation of atmospheric coherence time.} The second-order temporal coherence is calculated from the illumination intensity sequence measured by a high frame rate camera placed near the target. Using a model of second-order temporal coherence decaying exponentially with time delay, we fitted the atmospheric coherence time to approximately 27 ms.}
	\label{fig: g2_tau}
\addtocounter{Sfigure}{1}
\end{figure}

In the experiment, we have time bin $\Delta t=2$~ms, integration time $T=2$~s, count rate $r\approx 10^4$~Hz and $\Delta\tau$ is the atmospheric coherence time, which is measured to be about 27 ms as shown in Fig.~\ref{fig: g2_tau}. The value of $\Delta t$ satisfies the condition of the maximizing the SNR given by Eq. (\ref{snr3}), and it is calculated that SNR$\sim$10.

\end{document}